\documentclass[final,author year,3p,times,twocolumn]{elsarticle}
\usepackage{graphics}
\usepackage{amssymb}
\usepackage{bbm}

\usepackage{amsmath,textcomp,array}
\usepackage{bm}

\journal{New Astronomy}

\begin{document}

\begin{frontmatter}

\title{HACC: Simulating Sky Surveys on State-of-the-Art Supercomputing Architectures}
\author[ANLa,ANLb,KICP,CI]{Salman Habib}
\author[ANLa]{Adrian Pope}
\author[ANLc,ANLa]{Hal Finkel}
\author[ANLa,UC]{Nicholas Frontiere}
\author[ANLa,ANLb,KICP,CI]{Katrin Heitmann}
\author[LANL]{David Daniel}
\author[LANL]{Patricia Fasel}
\author[ANLc]{Vitali Morozov}
\author[Kitware]{George Zagaris}
\author[ANLb,CI]{Tom Peterka}
\author[ANLc,CI]{Venkatram Vishwanath}
\author[LBNL]{Zarija Luki\'c}
\author[NU]{Saba Sehrish}
\author[NU]{Wei-keng Liao}

\address[ANLa]{High Energy Physics Division, Argonne National
  Laboratory, Lemont, IL 60439, USA}
\address[ANLb]{Mathematics and Computer Science Division, Argonne
  National Laboratory, Lemont, IL 60439, USA}
\address[KICP]{Kavli Institute for Cosmological Physics, The University of
  Chicago, 5640 S. Ellis Ave., Chicago, IL 60637, USA}
\address[CI]{Computation Institute, The University of Chicago,
  Chicago, IL 60637}
\address[ANLc]{Argonne Leadership Computing Facility, Argonne National
  Laboratory, Lemont IL 60439}
\address[UC]{Department of Physics, University of Chicago, Chicago,
  IL 60637}
\address[LANL]{Computer, Computational, and Statistical Sciences
  Division, Los Alamos National Laboratory, Los Alamos, NM 87545} 
\address[Kitware]{Kitware, 28 Corporate Drive, Clifton Park, NY 12065}
\address[LBNL]{Computational Research Division, Lawrence Berkeley
  National Laboratory, Berkeley, CA 90095} 
\address[NU]{Department of Electrical Engineering and Computer
  Science, Northwestern University, 2145 Sheridan Road, Evanston, IL 60208}

\begin{abstract}
  Current and future surveys of large-scale cosmic structure are
  associated with a massive and complex datastream to study,
  characterize, and ultimately understand the physics behind the two
  major components of the `Dark Universe', dark energy and dark
  matter. In addition, the surveys also probe primordial perturbations
  and carry out fundamental measurements, such as determining the sum
  of neutrino masses. Large-scale simulations of structure formation
  in the Universe play a critical role in the interpretation of the
  data and extraction of the physics of interest. Just as survey
  instruments continue to grow in size and complexity, so do the
  supercomputers that enable these simulations. Here we report on HACC
  (Hardware/Hybrid Accelerated Cosmology Code), a recently developed
  and evolving cosmology N-body code framework, designed to run
  efficiently on diverse computing architectures and to scale to
  millions of cores and beyond. HACC can run on all current
  supercomputer architectures and supports a variety of programming
  models and algorithms. It has been demonstrated at scale on Cell-
  and GPU-accelerated systems, standard multi-core node clusters, and
  Blue Gene systems. HACC's design allows for ease of portability, and
  at the same time, high levels of sustained performance on the
  fastest supercomputers available. We present a description of the
  design philosophy of HACC, the underlying algorithms and code
  structure, and outline implementation details for several specific
  architectures. We show selected accuracy and performance results
  from some of the largest high resolution cosmological simulations so
  far performed, including benchmarks evolving more than 3.6 trillion
  particles.
 \end{abstract}


\end{frontmatter}

\section{Introduction}
\subsection{Sky Surveys and Computational Cosmology}

An unprecedented volume of observational data and information
regarding the distribution and properties of optical sources in the
Universe is becoming available from ongoing and future sky surveys,
both imaging and spectroscopic. These include BOSS (Baryon Oscillation
Spectroscopic Survey;~\citealt{boss}),
DES\footnote{http://www.darkenergysurvey.org/} (Dark Energy Survey;
\citealt{des}), KiDS (Kilo-Degree Survey; \citealt{dejong}),
SUMIRE\footnote{http://sumire.ipmu.jp/en/},
DESI\footnote{http://desi.lbl.gov/} (Dark Energy Spectroscopic
Instrument; \citealt{levi}), 4MOST (4-meter Multi-Object Spectroscopic
Telescope; \citealt{dejong2}), J-PAS (Javalambre-Physics of the
Accelerated Universe Astrophysical Survey; \citealt{benitez14}),
LSST\footnote{http://www.lsst.org/lsst/} (Large Synoptic Survey
Telescope;~\citealt{lsst}), Euclid~\citep{euclid}, and WFIRST
(Wide-Field InfraRed Survey Telescope; \citealt{spergel13}). Combined
with microwave background observations from the Planck
satellite\footnote{http://www.rssd.esa.int/index.php?project=Planck}
and ground-based telescopes, such as
ACT\footnote{http://www.princeton.edu/act/} (Atacama Cosmology
Telescope) and SPT\footnote{http://pole.uchicago.edu/} (South Pole
Telescope), the mining of these and other datasets, such as resulting
from new radio surveys, e.g., VLASS (Very Large Array Sky
Survey)\footnote{https://science.nrao.edu/science/surveys/vlass/vlass-white-papers}
and X-ray surveys, are expected to yield a host of cosmological and
astrophysical insights.

There are several foundational cosmological questions that the
datasets will directly address. Perhaps the most pressing is the
mysterious cause of the accelerated expansion of the Universe --
whether it is due to dark energy or a modification of general
relativity. In addition, the observations will also bear on the
ultimate nature of dark matter, provide information on the physics of
the early Universe by probing primordial fluctuations, and enhance our
knowledge of the neutrino family, the lightest known massive particles
in the Universe. Aside from these basic cosmological questions, the
survey data provides an enormous resource for attacking a very large
number of astrophysical problems related primarily to understanding
the formation of complex structure in the Universe.

In order to extract the full extent of information from these
remarkable surveys, a similar level of effort must be undertaken in
the realm of theory and modeling. To attain the necessary realism and
accuracy, sophisticated, large scale simulations of structure
formation must be carried out. These simulations address a large
variety of tasks: providing predictions for many different
cosmological models to solve the inverse problem related to
determining cosmological parameters, investigating astrophysical and
observational systematics that could mimic physics effects of the dark
sector, enabling careful calibration of errors (by providing precision
covariance estimates), testing, optimizing, and validating 
observational strategies with synthetic catalogs, and finally, 
exploring new and exciting ideas that could either explain puzzling
aspects of the observations (such as cosmic acceleration) or help to
motivate and design the implementation of new types of cosmological
probes.

\begin{figure}[t]
\begin{center}
\includegraphics[width=85mm]{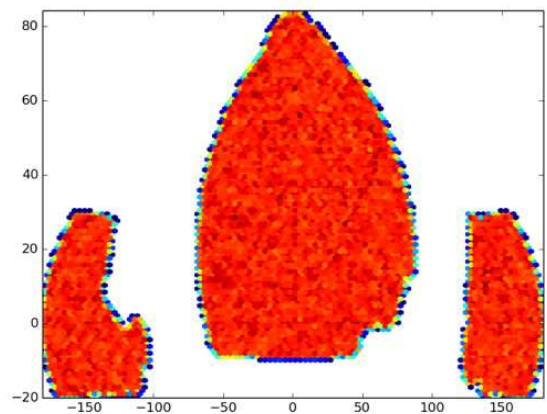}
\end{center}
\caption{Areal density of emission line galaxies from synthetic catalogs
  produced for the DESI experiment (\citealt{mw14}).  These catalogs
  are based on results from a high resolution, trillion particle HACC
  simulation performed on the Mira supercomputer. They are being used
  to test fiber assignment algorithms and to optimize DESI's footprint
  and survey strategy.}  
\label{syn_cat}
\end{figure}

The simulations have to be large enough in volume to cover the
observed Universe (or at least a large part of it) and at the same
time possess sufficient mass and force resolution (a spatial dynamic
range of roughly a million) to resolve objects down to the smallest
relevant scales. As one example, a recently constructed synthetic
galaxy and quasar catalog for DESI is shown in
Figure~\ref{syn_cat}. This catalog was based on a
1.1~trillion-particle HACC simulation, with a box-size of $\sim 4$~Gpc
(\citealt{mw14}).

The diverse uses of simulations outlined above also demand fast
turn-around times, as not one such simulation is required but,
eventually, many hundreds to hundreds of thousands. The simulation
codes, therefore, have to be capable of exploiting the largest
supercomputing platforms available today and into the future.

\subsection{Challenge of Future Supercomputing Architectures}  

Viewed as a single entity, the field of `modern' computational
cosmology (\citealt{klypin83}, \citealt{efstathiou85}) has largely
kept pace with the growth of computational power, but new challenges
will need to be faced over the next decade. This is due to the
failure of Dennard scaling~(\citealt{dennard}), which underlay the
success of Moore's Law for about two decades. As a consequence of the
fact that single-core clock rate and performance have stalled since
2004/5, the design of future microprocessors is branching into several
new directions, to overcome the related performance
bottlenecks~(\citealt{borkar}) -- the key constraint being set by
electrical power requirements. The resulting impact on large
supercomputers is already being seen; aside from the familiar large
multi-core processor clusters, two major approaches can be easily
identified.

The first approach is the large homogeneous system, built around a
`system on chip' (SoC) design or around many-core nodes, with
concurrencies in the multi-million range -- the IBM Blue Gene/Q is an
excellent example of such an approach; specific examples include
Sequoia (20~PFlops peak) at Lawrence Livermore National Laboratory and
Mira (10~PFlops peak) at Argonne National Laboratory, supporting up to
$6.3$ million-way concurrency (Sequoia; half of that on Mira). The
second route is to take a conventional cluster but to attach
computational accelerators to the CPU nodes. The accelerators can
range from the IBM Cell processor (as on Los Alamos National
Laboratory's Roadrunner, first to break the Petaflop barrier), to GPUs
as on Titan (27~PFlops peak) at Oak Ridge National Laboratory, to the
Intel Xeon Phi coprocessor as on Stampede (10~PFlops peak)
at the Texas Advanced Computing Center. Future evolution of
supercomputer systems will almost certainly involve further branching
in the space of possible architectures.

\begin{figure}[t]
  \centering \includegraphics[width=2.9in,angle=0]{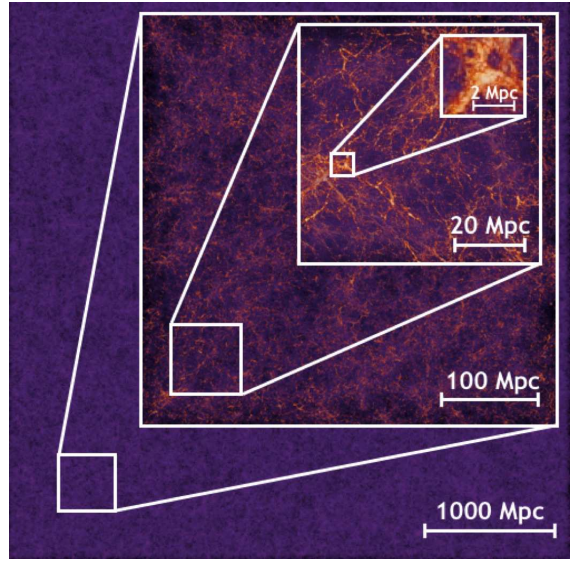}

  \caption{Zoom-in visualization of the density field in a 1.07
    trillion particle, $4.225~\hbox{Gpc}$ box-size simulation with
    HACC on 32 Blue Gene/Q racks. The force resolution is $6$~kpc and
    the particle mass, $m_p\sim 2.6\cdot 10^9$~M$_{\odot}$. The image,
    taken during a late stage of the evolution, illustrates the global
    spatial dynamic range covered by the simulation, $\sim 10^6$,
    although the finer details are not resolved in this
    visualization.}
\label{zoom}
\end{figure}

Because development of the supercomputing software environment is
likely to lag significantly behind the pace of hardware evolution,
it appears prudent, if not essential, to develop a code design
philosophy and implementation plan for cosmological simulations that
can be reconfigured relatively quickly for new architectures, has
support for multiple programming paradigms, and at the same time, can
extract high levels of sustained performance. With this challenge in
mind, we have designed HACC (Hardware/Hybrid Accelerated Cosmology
Codes), an N-body cosmology code framework that takes full advantage
of all available architectures. This paper presents an overview of
HACC `theory and practice' by covering the methods employed, as well
as describing important components of the implementation strategy. An
example of HACC's capabilities is shown in Figure~\ref{zoom}, a recent
large cosmological run on Mira, evolving more than 1 trillion
particles; this is the same simulation on which the results of
Figure~\ref{syn_cat} are based.
\noindent

\subsection{HACC Design Motivations and Implementation}

Cosmological simulations can be broadly divided into two classes:
gravity-only N-body simulations and `hydrodynamic' simulations that
incorporate gasdynamics and models of astrophysical processes. Since
gravity dominates at large scales, and dark matter outweighs baryons
by roughly a factor of five, N-body simulations are an essential
component in modeling the formation of structure. Several
post-processing strategies can be used to incorporate missing physics
in gravity-only codes, such as the halo occupation distribution (HOD)
approach (\citealt{kauffmann97}; \citealt{jing98}; \citealt{benson00};
\citealt{peacock00}; \citealt{seljak00}; \citealt{berlind02};
\citealt{zheng05}), Subhalo/Halo Abundance Matching (S/HAM)
(\citealt{vale04}; \citealt{conroy06}; \citealt{wetzel10};
\citealt{moster10}; \citealt{guo10}) or semi-analytic modeling (SAM)
(\citealt{white91}; \citealt{kauffmann93}; \citealt{cole94};
\citealt{somerville99}; \citealt{benson03}; \citealt{baugh06};
\citealt{benson10}) for adding galaxies to the simulations. Whenever a
more detailed understanding of the dynamics of baryons is required,
gasdynamic, thermal, and radiative processes (among others) must be
modeled, as well as processes such as star formation and local
feedback mechanisms (outflows, AGN/Sn feedback). A compact review of
cosmological simulation techniques and applications can be found in
\cite{dolag}. A concise review of phenomenological galaxy modeling is
given in \cite{baugh13}.

Carrying out a fully realistic first principles simulation program for
all aspects of modeling cosmological surveys will not be possible for
quite some time. The required gasdynamic simulations are very
expensive and there is considerable uncertainty in the modeling and
physics inputs. Progress is nevertheless possible by combining
high-resolution, large-volume N-body simulations and post-processing
inputs from simulations that include gas physics to build robust
phenomenological models. The parameters of these models would be
determined by a set of observational constraints; yet other
observations would then function as validation tests. The HACC
approach assumes this starting point as an initial design constraint,
but one that can be relaxed in the future.

The overall structure of HACC is based on the realization that a
large-scale computational framework must not only meet the challenges
of spatial dynamic range, mass resolution, accuracy, and throughput,
but, as already discussed, be cognizant of disruptive changes in
computational architectures. As an early validation of its design
philosophy, HACC was among the pioneering applications proven on the
heterogeneous architecture of Roadrunner~(\citealt{hacc1,hacc2}), the
first computer to break the petaflop barrier. With its
multi-algorithmic structure, HACC allows the coupling of MPI with a
variety of local programming models -- MPI+`X' -- to readily adapt to
different platforms. Currently, HACC is implemented on conventional
and Cell/GPU-accelerated clusters, on the Blue Gene/Q architecture
(\citealt{hacc3}), and has been run on prototype Intel Xeon Phi
hardware. HACC is the first, and currently the only large-scale
cosmology code suite world-wide, that has been demonstrated to run at
full scale on {\em all} available supercomputer architectures.

Another key aspect of the HACC code suite is an inbuilt capability for
fast ``on the fly'' or {\em in situ} data analysis. Because the raw
data from each run can easily be at the petabyte (PB) scale or larger,
it is essential that data compression and data analysis be maximized
to the extent possible, before the code output is dumped to the file
system. In order to comply with storage and data transfer bandwidth
limitations, the amount of data reduction required is roughly two
orders of magnitude. HACC's {\em in situ} data analysis system is
designed to incorporate a number of parallel tools such as tools for
generating clustering statistics (power spectra, correlation
functions), tessellation-based density estimators, a fast halo
finder~\citep{woodring11} with an associated merger tree capability,
real-time visualization, etc.

The HACC framework has been used to generate a number of large
simulations to carry out a variety of scientific projects. These
include a suite of 64~billion particle runs for predicting the baryon
acoustic oscillation signal in the quasar Ly-$\alpha$ forest, as
observed by BOSS~(\citealt{hacclya}), a high-statistics study of
galaxy group and cluster profiles, to better establish the halo
concentration-mass relation~(\citealt{bhattacharya11_b}), tests of a
new matter power spectrum emulator~(\citealt{heitmann14}), and a study
of the effect of neutrino mass and dynamical dark energy on the matter
power spectrum~(\citealt{upadhye}). Results from other simulation runs
will be available shortly. These range from simulation campaigns in
the $\sim 70$~billion particle range per simulation, e.g., simulations
for determining sampling covariance for BOSS~(\citealt{sunayama}), to
individual simulations in the $\sim 500-1000$ billion particle
class. As an example of the latter, a recently completed simulation on
Titan used 550 billion particles to evolve a 1.3~Gpc periodic box,
with mass resolution, $m_p=1.48\cdot 10^8$~M$_\odot$, and force
smoothing scale of $\sim 3.2$~kpc~(\citealt{heitmann14b}).

The paper is organized as follows. In Section~\ref{sec:feat} we
introduce the basic HACC design and algorithms, focusing on how
portability and scaling on very large supercomputers is achieved,
including discussions of practical matters such as memory management
and parallel I/O. A significant aspect of supercomputer architecture
is diversity at the level of the computational nodes. As mentioned
above, the HACC design adopts different short-range solvers for
different nodal architectures, and this feature is discussed
separately in Section~\ref{sec:specs}. Section~\ref{sec:verif}
presents some results from our extensive code verification program,
showcasing a comparison test with {\sc Gadget-2}~\citep{gadget2}, one
of the most widely used cosmology codes today. The {\em in situ} analysis
tool suite is covered in Section~\ref{sec:tools}, and selected
performance results are given in Section~\ref{sec:perf}.  We conclude
in Section~\ref{sec:conc} with a recap and a discussion of future
evolution paths for HACC.

\section{General Features of the HACC Code Framework}
\label{sec:feat}
\subsection{N-body Algorithms}

In the standard model of cosmology, structure formation at large
scales is described by the gravitational Vlasov-Poisson equation
(\citealt{peebles}), a 6-D partial differential equation for the
Liouville flow (\ref{le}) of the one-particle phase space
distribution, arising from the non-relativistic limit of the
Vlasov-Einstein set of equations,
\begin{equation}
\partial_t f({\mathbf{x}},
{\mathbf{p}})+\dot{\mathbf{x}}\cdot\partial_{\mathbf{x}}
f({\mathbf{x}}, {\mathbf{p}}) -
\nabla\phi\cdot\partial_{\mathbf{p}}f({\mathbf{x}}, 
{\mathbf{p}})=0,\label{le}
\end{equation}
where the Poisson equation encodes the self-consistency of the
evolution: 
\begin{equation}
\nabla^2\phi({\mathbf{x}})= 4\pi
Ga^2(t)\Omega_m\delta_m({\mathbf{x}})\rho_c.
\label{pe}
\end{equation}
The expansion history of the Universe is given by the time-dependence
of the scale factor $a(t)$ governed by the specifics of the
cosmological model, the Hubble parameter, $H=\dot{a}/a$, $G$ is
Newton's constant, $\rho_c$ is the critical density, $\Omega_m$, the
average mass density as a fraction of $\rho_c$, $\rho_m({\mathbf{x}})$
is the local mass density, and $\delta_m({\mathbf{x}})$ is the
dimensionless density contrast,
\begin{equation}
\rho_c=3H^2/8\pi G,~~~
\delta_m({\mathbf{x}})=(\rho_m({\mathbf{x}})-
\langle\rho_m\rangle)/\langle\rho_m\rangle,
\label{defs1}
\end{equation}
\begin{equation} 
{\mathbf{p}}=a^2(t) \dot{\mathbf{x}},
~~~\rho_m({\mathbf{x}})= a(t)^{-3}m\int
d^3{\mathbf{p}}f({\mathbf{x}}, {\mathbf{p}}). 
\label{defs2}
\end{equation}
In general, the Vlasov-Poisson equation is computationally very
difficult to solve as a partial differential equation because of its
high dimensionality and the development of nonlinear structure --
including complex multistreaming -- on ever finer scales, driven by
the gravitational Jeans instability. Consequently, N-body methods,
using tracer particles to sample $f({\mathbf{x}}, {\mathbf{p}})$ are
used; the particles follow Newton's equations, with the forces given
by the gradient of the scalar potential
$\phi(\mathbf{x})$~(Cf. \citealt{braun_hepp77}; for an early
comparison of direct and particle methods in a non-gravitational
context, see \citealt{denavit71}).

The cosmological N-body problem is characterized by a very large value
of the number of interacting particles and a very large spatial
dynamic range. If one wishes to track billions of galaxy-hosting halos
at a reasonable mass resolution, then hundreds of billions to
trillions of tracer particles are required. Since gravity cannot be
shielded, this obviously precludes the use of brute-force direct
particle-particle algorithms for the particle force
computation. Popular alternatives include pure particle-based methods
(tree codes) or multi-scale grid-based methods (AMR codes), or hybrids
of the two (TreePM, particle-particle particle-mesh, P$^3$M). It is
not our purpose here to go into many details of the algorithms
and their implementations; good coverage of the background material
can be found in \cite{barnes86}, \cite{hockney}, \cite{warren93},
\cite{pfalzner}, \cite{dubinski}, \cite{gadget2}, and \cite{dolag}.

The HACC design approach acknowledges that, as a general rule,
particle and grid-based methods both have their limitations. For
physics, algorithmic, and data structure reasons, grid-based techniques
are better suited to larger (`smooth') lengthscales, with particle
methods possessing the opposite property. This suggests that higher
levels of code organization should be grid-based, interacting with
particle information at a lower level of the computational hierarchy.

Following this central idea, HACC uses a hybrid parallel algorithmic
structure, splitting the gravitational force calculation into a
specially designed grid-based long/medium range spectral particle-mesh
(PM) component that is retained on all computational architectures,
and an architecture-tunable particle-based short/close-range
solver. The spectral PM component can be viewed as an upper layer that
is implemented using C++/C/MPI, essentially independent of the target
architecture, whereas, the bottom or node level is where the
short/close-range solvers reside. These are chosen and optimized
depending on the target architecture and use different local
programming models as appropriate. Of the 6 orders of magnitude
required for the spatial dynamic range for the force solver, the grid
is responsible for 4 orders of magnitude, while the particle methods
handle the critical 2 orders of magnitude at the shortest scales where
particle clustering is maximal and the bulk of the time-stepping
computation takes place.

The short-range solvers can employ direct particle-particle
interactions, i.e., a P$^3$M algorithm~\citep{hockney}, as on some
accelerated systems, or use both tree and particle-particle methods as
on the IBM Blue Gene/Q (`PPTreePM' with a recursive coordinate
bisection (RCB) tree). As two extreme cases, in non-accelerated
systems, the tree solver provides very good performance but has some
complexity in the data structure, whereas for accelerated systems, the
local $N^2$ approach is more compute-intensive but has a very simple
data structure, better-suited for computational accelerators such as
Cells and GPUs. The availability of multiple algorithms within the
HACC framework also allows us to carry out careful error analyses, for
example, the P$^3$M and the PPTreePM versions agree to within $0.1\%$
for the nonlinear power spectrum test in the code comparison suite of
~\cite{heitmann05} (see Section~\ref{sec:verif} for more
details). This level of error control easily meets the minimal
requirements set by the increased statistical power of next-generation
survey observations.

In the following we first describe the long/medium-range force solver
employed by HACC. As mentioned above, this solver remains unchanged
across all architectures. After doing this, we provide details of the
architecture-specific short-range solvers, and the sub-cycled
time-stepping scheme used by the code suite.

\subsection{Particle Overloading}
\label{sec:overload}

An important aspect of large-volume cosmological simulations is that
the density distribution is very highly clustered, with an overall
topology descriptively referred to as the ``cosmic web''. The
clustering is such that the maximum distance moved by a particle is
roughly 30~Mpc, very much smaller than the overall scale of the
simulation box ($\sim$Gpc). With a 3-D domain decomposition, each
(non-cubic) nodal volume (MPI rank) is roughly of linear size
50-500~Mpc, depending on the simulation run size. The idea behind
particle overloading is to `overload' the node with particles
belonging also to a zone of size roughly 3-10~Mpc extending out from
the nominal spatial boundary for that node (so-called ``passive''
particles). Note that copies of these particles -- essentially a
replicated particle cache, roughly analogous to ghost zones in PDE
solvers -- will also be held by other processors, in only one of which
will they be ``active'', hence the use of the term
`overloading'. Because more than one copy of these particles is held
by neighboring domains, overloading is not the same as the guard zone
conventionally used to reduce communication in particle codes.

The point of having this particle cache is two-fold. First, for a
number of time steps no particle communication across nodes is
required. Additionally, the cached particle `skin' allows particle
deposition and force interpolation for the spectral particle-mesh
method to be done using information entirely local to a node, thus
grid communication is also reduced. The particle cache is refreshed
(replacement of passive particles in each computational domain by
active particles from neighboring domains) at some given number of
time steps. This involves only nearest neighbor communication and the
penalty is a trivial fraction of the time spent in a single global
time step. The second advantage of overloading is that when a
short-range solver is added to each computational domain, no
communication infrastructure is associated with this step. Thus, in
principle, short-range solvers can be developed completely
independently at the node level, and then inserted into the code as
desired. Consequently, HACC's weak scaling is purely a function of the
properties of the long-range solver.

The overloaded PM solver is formally exact -- each node sends its
local density field (computed with active particles only) to the
global spectral Poisson solver, which then returns the force for both
active and passive particles to each node. The short-range force
calculation for passive particles is computed in the same way as for
the active particles, except that passive particles, which are closer
to the outer domain boundary than the short-range force-matching
scale, $r_s$ (defined in the following Section), do not have their
short-range forces computed, and are subject to only long-range
forces. This avoids force anisotropy near the overloaded zone
boundary, at the expense of a force error on the passive particles
that are close to the edge of the boundary. Since the role of the
passive particles is primarily to provide a buffer `boundary
condition' zone for the active particles near the nodal domain's
active zone boundary, consequences of this error are easy to control.

Overloading has two associated costs: (i) loss of memory efficiency
because of the overloading zone, and (ii) the just-discussed numerical
error for the short-range force that slowly leaks in from the edge of
the outer (passive particle) domain boundary. In cosmology
applications, the memory inefficiency can be tolerated to the point
where the memory in passive and active particles is roughly equal, but
this is not a limitation in the majority of the cases of interest. The
second problem is easily mitigated by balancing the boundary thickness
against the frequency of the particle cache refresh, ``recycling'' the
passive particles after some finite number of time-steps, chosen to
be such that the refresh time is smaller than the error diffusion time:
each domain gets rid of all of its passive particles and then
refreshes the passive zone with (high accuracy) active particles
exchanged from nearest-neighbor domains.

\subsection{The Long-Range Force} 
\label{sec:lrange}

Conventional particle-based codes employ a combination of spatial and
spectral techniques. The Cloud-in-Cell (CIC) scheme used for particle
deposition is an example of a real space operation, whereas the Fast
Fourier Transform (FFT)-based Poisson solver is a $k$-space or
spectral operation. The spatial operations in a typical particle code
are the particle deposition, the force interpolation, and the
finite-differences for computing field derivatives. The spectral
operations include the influence (or pseudo-Green) function FFT solve,
digital filtering, and spectral differentiation techniques.  Spatial
techniques are often less flexible and more tedious to implement than
their spectral counterparts. Also, higher-order spatial operations can
be complicated and lead to messy and dense communication patterns
(e.g., indirection).

Accurate P$^3$M codes are usually run with Triangle-Shaped-Cloud
(TSC), a high-order spatial deposition/filtering scheme, as well as
with high-order spatial differencing templates. In terms of grid
units, the CIC deposition kernel filters roughly at the level of two
grid cells or less with a large amount of associated ``anisotropy
noise''; TSC filters at about three grid cells with much reduced
noise. The resulting long-range/short-range force matching is usually
done at four/five grid cells or so (but can be as small as three grid
cells). TreePM codes can move the matching point to a larger number of
grid cells because of the inherent speed of the tree algorithm, so
they can continue to use CIC deposition (with some additional Gaussian
filtering). HACC allows the use of both P$^3$M and TreePM algorithms,
using a shorter matching scale than most TreePM codes. One advantage
of the shorter matching scale is that a low-order polynomial
expression can be used in the force representation, which greatly
speeds up evaluation of the force kernel.

\begin{figure}[t]
\begin{center}
\includegraphics[width=85mm,angle=0]{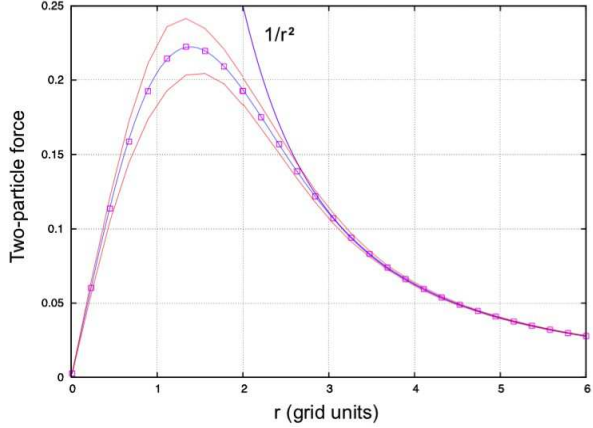}
\end{center}
\caption{Orientation-averaged force between two particles as a
  function of separation, using digital filtering and spectral
  differentiation with CIC deposition as implemented in HACC. The
  match to the exact $1/r^2$ behavior is tuned to occur at a
  separation of three grid cells, which sets the force-matching scale,
  $r_s$. The low level of the force anisotropy noise is shown by the
  bracketing red lines representing the 1-$\sigma$ deviation. The
  solid curve is the fitting formula of Eq.~\ref{fitform}.}
\label{ppforce}
\end{figure}

The HACC long-range force solver uses only CIC mass deposition and
force interpolation with a Gauss-sinc spectral filter to mimic
TSC-like behavior. In addition, a fourth-order Super-Lanczos spectral
differentiator~(\citealt{hamming}) is used, along with a sixth-order
influence function. This approach allows the data motion to be
simplified as no complicated spatial differentiation is needed. Also,
by moving more of the Poisson-solve to the spectral domain, the
inherent flexibility of Fourier space can be exploited. For example,
the filtering is flexible and tunable, allowing careful force matching
at only three grid cells.  Figure~\ref{ppforce} shows the
force-matching with the spectral techniques using a pair of test
particles. In this test, multiple realizations of particle pairs were
taken at fixed distances, but with random orientations of the
separation vector in order to sample the anisotropy imposed by the
grid-based calculation of the force.

The solution of the Poisson equation in HACC's long range force-solver
is carried out using large FFTs; the corresponding spectral
representation of the inverse (discrete) Laplacian is the influence
function. HACC employs the following three-dimensional, sixth-order,
periodic, influence function:
\begin{eqnarray}
  G_6({\bf k})&=&\frac{45}{128}\Delta^2\left[\sum_i\cos\left(\frac{2\pi
    k_i\Delta}{L}\right)\right.\nonumber\\  
 &-& \frac{5}{64}\sum_i\cos\left(\frac{4\pi
      k_i\Delta}{L}\right)\nonumber\\ 
 &+& \left.\frac{1}{1024} \sum_i\cos\left(\frac{8\pi
       k_i\Delta}{L}\right) - \frac{2835}{1024}\right]^{-1}, 
\label{greenf}
\end{eqnarray}
where the sum is over the three spatial dimensions, $\Delta$ is the
grid spacing, and $L$ the physical box size.

As mentioned earlier, the CIC-deposited density field is spectrally
filtered using a sinc-Gaussian filter:
\begin{equation}
S({\bf k})=\exp{(-\left|{\bf k}\right|^2\sigma^2/4)}\left[(2/{\bf
    k}\Delta)\sin({\bf k}\Delta/2)\right]^{n_s}. 
\label{filter}
\end{equation}
The aim of the filter is to reduce the anisotropy noise as well as
control the matching scale where the short-range and long-range forces
are matched; the nominal choices in HACC are $\sigma=0.8$ and
$n_s=3$. This filtering reduces the anisotropy ``noise'' of the basic
CIC scheme by better than an order of magnitude, and allows for the
use of the higher-order spectral differencing scheme.

Instead of solving for the scalar potential, and then using a spatial
stencil-based differentiation, HACC uses spectral differentiation
within the Poisson-solve itself, using fourth-order spectral Lanczos
derivatives, as previously mentioned. The (one-dimensional)
fourth-order Super-Lanczos spectral differentiation for a function,
$f$, given at a discrete set of points is
\begin{eqnarray}
&&\left.\frac{\Delta f}{\Delta x}\right|_{4}\simeq \nonumber\\
&&\frac{4}{3} \sum_{j=-N+1}^{N}C_j{\rm e}^{(2\pi j x/L)} i\frac{2\pi j\Delta}{L}
\frac{\sin\left(2\pi j\Delta/L\right)}{2\pi j\Delta/L} \nonumber \\
&&-\frac{1}{6} \sum_{j=-N+1}^{N}C_j{\rm e}^{(2\pi j x/L)} i\frac{2\pi j\Delta}{L}
\frac{\sin\left(4\pi j\Delta/L\right)}{2\pi j\Delta/L}, 
\label{sldiff}
\end{eqnarray}
where $C_j$ are coefficients in the Fourier expansion of $f$. 

The ``Poisson-solve'' in the HACC code is the composition of all the
kernels above in one single Fourier transform. Note that each
component of the field gradient requires an independent FFT. This
entails some extra work, but is a very small fraction of the total
force computation, the bulk of which is dominated by the short-range
solver.

An efficient and scalable parallel Fast Fourier Transform (FFT) is an
essential component of HACC's design, and determines its weak scaling
properties. Although parallel FFT libraries are available, HACC uses
its own portable parallel FFT implementation optimized for low memory
overhead and high performance. Since slab-decomposed parallel FFTs are
not scalable beyond a certain point (restricted to
$N_{rank}<N_{FFT}$), the HACC FFT implementation uses data
partitioning across a two-dimensional subgrid, allowing
$N_{rank}<N^2_{FFT}$, where $N_{rank}$ is the number of MPI ranks and
$N_{FFT}$ is the linear size of the 3-D array. The resulting scalable
performance is sufficient for use in any supercomputer in the
foreseeable future~(\citealt{hacc3}).

\begin{figure}[t]
\begin{center}
\includegraphics[width=65mm,angle=0]{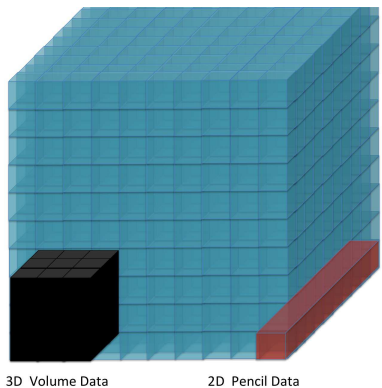}
\end{center}
\caption{Data allocations for the force calculation. A
  three-dimensional spatial domain decomposition is used for for the
  force-solver, while a two-dimensional pencil structure is used for
  the FFT. Therefore, a reallocation of memory between the two data
  structures is required when carrying out either step in the
  computation.} 
\label{pencil}
\end{figure}

The implementation consists of a data partitioning algorithm which
allows an FFT to be taken in each dimension separately. The data
structure of the computing nodes prior to the FFT is such as to divide
the total space into regular three-dimensional domains. Therefore, to
employ a two-dimensionally decomposed FFT, the distribution code
reallocates the data from small `cubes', where each cube represents
the data of one MPI rank, to thin two-dimensional `pencil' shapes, as
depicted schematically in Figure~\ref{pencil}.

Once the distribution code has formed the pencil data decomposition, a
one-dimensional FFT can be taken along the long dimension of the
pencil. Moreover, the same distribution algorithm is employed to carry
out the remaining two transforms by redistributing the domain into
pencils along those respective dimensions. The transposition and FFT
steps are overlapped and pipelined, with a reduction in communication
hotspots in the interconnect. Lastly, the dataset is returned to the
three-dimensional decomposition, but now in the spectral
domain. Pairwise communication is employed to redistribute the data,
and has proven to scale well in our larger simulations. A
demonstration of this is provided by the Blue Gene/Q sytems, where we
have run on up to $\sim1.5$ million MPI ranks (\citealt{hacc3}). As
the grid size is increased on a given number of processors, the
communication efficiency (i.e., the fraction of time spent
communicating data between processors), remains unchanged. This is an
important validation of our implementation design, as the
communication cost of the algorithm must not outpace the increase in
local computation performance when scaling up in size. Further details
of the parallel FFT implementation will be presented elsewhere.

\subsection{The Short-Range Force}
\label{srf}

The total force on a particle is given by the vector sum of two
components: the long-range force and the short-range force. At
distances greater than the force-matching scale, only the long-range
force is needed (at these scales, the (filtered) PM calculation is an
excellent approximation to the desired Newtonian limit, see
Figure~\ref{ppforce}). At distances less than the force-matching
scale, $r_s$, the short-range force is given by subtracting the
residual filtered grid force from the exact Newtonian force.

To find the residual filtered PM force, we compute it numerically
using a pair of test particles (since in our case no analytic
expression is available), evaluating the force at many different
distances at a large number of random orientations. The results are
fit to an expression that has the correct asymptotic behaviors at
small and large separation distances (Cf.~\citealt{dubinski}). We used
the particular form:
\begin{eqnarray}
f_{grid}(r)&=&\frac{1}{r^2}\tanh(br)-\frac{b}{r}\frac{1}{\cosh^2(br)}
\nonumber\\  
&+&cr(1+dr^2)\exp(-dr^2)\nonumber\\
&+&e(1+fr^2+gr^4+lr^6)\exp(-hr^2),
\label{fitform}
\end{eqnarray}
which, with $b=0.72$, $c=0.01$, $d=0.27$, $e=0.0001$, $f=360$,
$g=100$, $h=0.67$, and $l=17$, provides an excellent match to the data
from the test code (Figure~\ref{ppforce}), with errors much below
$0.1\%$. At very small distance scales, the gravitational force must
be softened, and this can be implemented using Plummer or spline
kernels (see, e.g., \citealt{dehnen01}).

The force expression, Eq.~(\ref{fitform}), is complex and to implement
faster force evaluations one can either employ look-ups based on
interpolation or a simpler polynomial expression. The communication
penalty of look-ups can be quite high, whereas an extended dynamic
range is difficult to fit with sufficiently low-order polynomials. In
our case, the choice of a short matching scale, $r_s$, enables the use
of a fifth-order polynomial approximation, which can be vectorized for
high performance. 

Depending on the target architecture, HACC uses two different
short-range solvers, one based on a tree algorithm, the other based on
a direct particle-particle interaction (P$^3$M). Tree methods are
employed on non-accelerated systems, while both P$^3$M and tree
methods can be used on accelerated systems. HACC uses an RCB tree in
conjunction with a highly-tuned short-range polynomial force
kernel. (An oct-tree implementation also exists, but is not the
current production version.) The implementation of the RCB tree,
although not the force evaluation scheme, generally follows the
discussion in~\cite{gafton}.  (Multiple RCB trees are used per rank to
enhance parallelism, as described in Section~\ref{rcb}.) Two core
principles underlie the high performance of the RCB tree's design
(\citealt{hacc3}).

\begin{figure}[t]
\begin{center}
\includegraphics[width=75mm,angle=0]{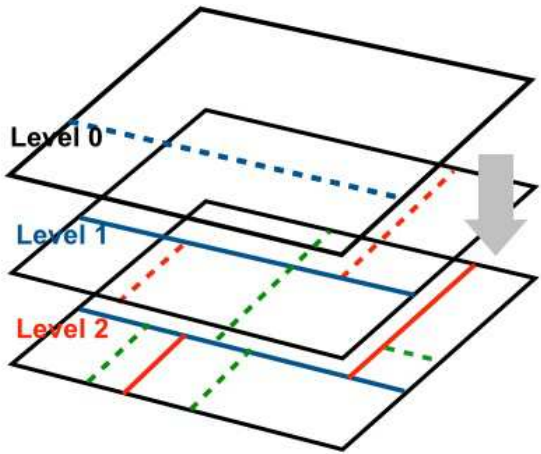}
\end{center}
\caption{Notional sketch of an RCB tree spatial domain splitting for a
  2-D particle distribution, following \cite{gafton}. The cut-lines
  (planes in 3-D) go through the particle distribution center of mass
  in a cell -- rather than the median -- and are perpendicular to the
  longest cell side. The criterion for the maximal level of the tree
  depth depends on the details of the computational architecture (see
  text).}
\label{rcb}
\end{figure}

{\em Spatial Locality.} The RCB tree is built by recursively dividing
particles into two groups. The dividing line is placed at the center
of mass coordinate perpendicular to the longest side of the box
(Figure~\ref{rcb}). Once the line of division is chosen, the particles
are partitioned such that particles in each group occupy disjoint
memory buffers. Local forces are then computed one leaf node at a
time. The net result is that the particle data exhibits a high degree
of spatial locality after the tree build; because the computation of
the short-range force on the particles in any given leaf node, by
construction, deals with particles only in nearby leaf nodes, the
cache miss rate during the force computation is very low.

{\em Walk Minimization.} In a traditional tree code, an interaction
list is built and evaluated for each particle. While the interaction
list size scales only logarithmically with the total number of
particles (hence the overall ${\mathcal{O}}(N\log N)$ complexity), the
tree walk necessary to build the interaction list is a relatively slow
operation. This is because it involves the evaluation of complex
conditional statements and requires ``pointer chasing'' operations. A
direct $N^2$ force calculation scales poorly as $N$ grows, but for a
small number of particles, a thoughtfully-constructed kernel can still
finish the computation in a small number of cycles. The RCB tree
exploits our highly-tuned short-range force kernels to decrease the
overall force evaluation time by shifting workload away from the slow
tree-walking and into the force kernel. Up to a point, doing this
actually speeds up the overall calculation: the time spent in the
force kernel goes up but the walk time decreases faster. Obviously, at
some point this breaks down, but on many systems, tens or hundreds of
particles can be in each leaf node before the crossover is reached. We
point out that the force kernel is generally more efficient as the
size of the interaction list grows: the relative loop overhead is
smaller, and more of the computation can be done using unrolled
vectorized code.

In addition to the performance benefits of grouping multiple particles
in each leaf node, doing so also increases the accuracy of the
resulting force calculation: The local force is dominated by nearby
particles, and as more particles are retained in each leaf node, more
of the force from those nearby particles is calculated exactly. In
highly-clustered regions (with very many nearby particles), the
accuracy can increase by several orders of magnitude when keeping over
100 particles per leaf node.

The P$^3$M implementation within HACC follows the standard method of
building a chaining mesh to control the number of particle-particle
interactions (\citealt{hockney}). This algorithm is used within HACC
when working with accelerated systems. We defer further details
regarding the architecture-specific implementation of the short-range
force to Section~\ref{sec:specs}, where the different alternatives are
covered separately.

\subsection{Time-Stepping}
\label{time-stepper}

The time-stepping in HACC is based on the widely employed symplectic
scheme, as used, e.g., in the IMPACT code (\citealt{impact}), the
forerunner of MC$^2$~(Mesh-based Cosmology Code, see, e.g.,
\citealt{heitmann05}), in turn the PM precursor of HACC. The basic
idea here is not to finite-difference the equations of motion, but to
view evolution as a symplectic map on phase space. Symplectic
integration in HACC approximates the full evolution to second order in
the time-step by composing elementary maps using the
Campbell-Baker-Hausdorff series expansion (\citealt{sternberg}). In PM
mode, the elementary maps are the `stream' and `kick' maps
$M_1=\exp(-t\hat{H}_1)$ and $M_2=\exp(-t\hat{H}_2)$ corresponding to
the free particle (kinetic) piece and the one-particle effective
potential in the Hamiltonian, respectively. In the stream map, the
particle position is drifted using its known velocity, which remains
unchanged; in the kick map, the velocity is updated using the force
evaluation, while the position remains unchanged. A symmetric
`split-operator' symplectic step $M_1(t/2)M_2(t)M_1(t/2)$ is termed
SKS (stream-kick-stream); a KSK step is another way to implement a
second-order symplectic integrator. (In the presence of explicitly
time-dependent Hamiltonan pieces, the map evaluations have to be
implemented at the correct times to maintain second-order accuracy.)

In the presence of both short and long-range forces, we split the
Hamiltonian into two parts, $H_1=H_{sr} + H_{lr}$ where $H_{sr}$
contains the kinetic and particle-particle force interaction (with an
associated map $M_{sr}$), whereas, $H_2=H_{lr}$ is just the long range
force, corresponding to the map $M_{lr}$. Since the long range force
varies relatively slowly, we construct a single time-step map by
subcycling $M_{sr}$:
\begin{equation}
M_{full}(t)=M_{lr}(t/2)(M_{sr}(t/n_c))^{n_c}M_{lr}(t/2),
\label{tstep}
\end{equation}
the total map $M_{sr}$ being a usual second-order symplectic
integrator. This corresponds to a KSK step, where the S is not an
exact stream step as in the PM case, but has enough $M_{sr}$ steps
composed together to obtain the required accuracy. For typical
problems the number, $n_c$, of short time steps for each long time
step will range between 3-10, depending on accuracy requirements.

Because the late-time distribution of particles is highly clustered,
there can be a substantial advantage in using different (synchronized)
local time steps down to the single-particle level. Although HACC is
currently designed for a regime where extreme dynamic range is not
needed, as in treating the innermost part of galaxy halos or in
tracking orbits around black holes -- where this advantage is most
felt (e.g., \citealt{power2003}) -- the automatic density information
available in the short-range force solvers is used to enable
multi-level time-stepping, resulting in speed-ups by a factor of 2-3,
with only small effects on the accuracy. More on this topic can be
found in Section~\ref{sec:specs}.

\subsection{Code Units}
\label{units}

HACC uses comoving coordinates for positions and velocities. The
actual internal representation of all variables is in the
dimensionless form:
\begin{equation}
{\bf x}\equiv
x_0{\bf{\tilde{x}}},~~t\equiv\tilde{t}/H_0,~~\rho\equiv\tilde{\rho}\rho_b, 
\label{scaling}
\end{equation}
where the fundamental scaling length, $x_0$, is the length of a single
grid cell of the long-range force PM-solver, $L/(N_g-1)$, where $L$ is
the box-size and $N_g$ is the number of grid points in a single
dimension, $H_0$ is the current value of the Hubble parameter, and the
background mass density, $\rho_b=3H_0^2\Omega_0/(8\pi G a^3)$. HACC
uses powers of the scale factor, $a(t)^{\alpha}$, as the actual
evolution variable, with a nominal default value of $\alpha=1$;
time-stepping is performed using the variable $y=a^{\alpha}$,
$d/d\tilde{t}=(\alpha y H/H_0)d/dy$.

\subsection{Memory Management}

Besides easy portability between different architectures, another very
important feature of HACC is its highly optimized memory
footprint. Pushing the simulation limits in large-scale structure
formation problems means running simulations with as many particles as
possible, and this often implies running as close as possible to the
memory limit of the machine. As a result, memory
fragmentation\footnote{Memory fragmentation refers to the condition
  where small allocations dispersed throughout the memory space leave
  no large contiguous chunks free, even though the total amount of
  free memory may be large.} becomes a serious problem. To make
matters worse, HACC is required to allocate and free different data
structures during different parts of each time step because there is
not enough available memory to hold all such structures at the same
time. Furthermore, many of these data structures, such as the RCB tree
used for the short-range force calculation, have sizes that change
dynamically with each new time step. This, combined with other
allocations from the MPI implementation, message printing, file I/O,
etc. with lifetimes that might outlast a time-step phase
(e.g. long-range force computation, short-range force computation, {\em
  in situ} analysis), is a recipe for fatal memory fragmentation
problems -- problems that we actually encountered on the Blue Gene/Q
systems.

To mitigate this difficulty we have implemented a specialized pool
allocator called Bigchunk. This allocator grabs a large chunk of
memory, and then distributes it to various other subsystems. During
the first time step, Bigchunk acts only as a wrapper of the system's
memory allocator, except that it keeps track of the total amount of
memory used during each phase of the time step. Before the second time
step begins, Bigchunk allocates an amount of memory equal to the
maximum used during any phase of the previous time step plus some
safety factor. Subsequent allocations are satisfied using memory from
the `big chunk', and all such memory is internally marked as free after
the completion of each time-step phase. This en-masse deallocation,
the only kind of deallocation supported by Bigchunk, allows for an
implementation that has minimal overhead, and the time it takes to
allocate memory from Bigchunk is very small compared to the speed of
the system allocator. Because the Bigchunk memory is not released back
to the system, memory fragmentation no longer fatally affects the
ability of the time-step phases to allocate their necessarily-large
data structures.

The persistent state information in the simulation is carried by the
particle attributes. While the number of particles in each MPI rank's
(overloaded) spatial sub-volume is similar, structure formation
implies some variance. Once the overload cache is filled, the (total)
number of particles on a rank is fixed until the cache is emptied and
refreshed, at which point the number of particles on each rank can
change. In order to avoid memory fragmentation with persistent
information we must anticipate the maximum number of particles any
rank will need during the course of a simulation run and allocate that
amount of memory for particle information at the start, before any
other significant memory allocation occurs. We estimate the maximum
number of particles by choosing a maximum representative volume from
which bulk motion could move all of the particles into a rank's
(overloaded) sub-volume and multiply that volume by the average
particle density. The actual memory allocations are monitored during
runtime, allowing for adjustments to be made while the code is
running. In practice for the sub-volumes with side lengths that are at
least several tens of Mpc this results in allocating memory for an
additional skin of particles that is 6-10~Mpc thick. HACC prints a
memory diagnostic at each time step to indicate the extrema of
particle memory usage across all ranks, and the amount of extra memory
can be adjusted when restarting the code if the initial estimate
appears to be insufficient for later times. In addition to the space
required for the actual particle information, we also allocate an
array of integers to represent a permuted ordering of the particles
and a scratch array large enough to hold any single particle
attribute. These enable out-of-place re-ordering of the particle
information one attribute at a time without additional memory
allocation.

\subsection{I/O Strategy}

A plurality of I/O strategies are needed for different use cases,
machine architectures, and data sizes because no single I/O strategy
works well under all of these conditions. Our three main approaches
are described below.

\textit{One file per process.} Using one output file per process
(i.e. MPI rank) is the simplest I/O strategy, and continues to provide
the best write bandwidth compared to any other strategy. Because every
process writes into a separate file, after file creation, there is no
locking or synchronization needed in between processes. Unfortunately,
while simple and portable, one file per process only works for a
modest number of processes (typically less than 10,000). No file
system can manage hundreds of thousands of files that would result
from checkpointing a large-scale run with one file per
process. Additionally, as a practical matter, managing hundreds of
thousands of files is cumbersome and error-prone. Finally, even when
the number of files is reasonable, reading the stored data back into a
different number of processes than used to write the data requires
redistribution. This happens when the output is used for analysis on a
smaller cluster (or machine partition), or for visualization. In such
cases, the required reshuffling of all of the data in memory is
equivalent to the aggregation done by more complex collective I/O
strategies and cancels out the simplicity of the one file per process
approach. For improved scalability and flexibility, HACC supports the
following additional I/O strategies.

\begin{figure}[t]
  \centering 
  \includegraphics[width=2.8in]{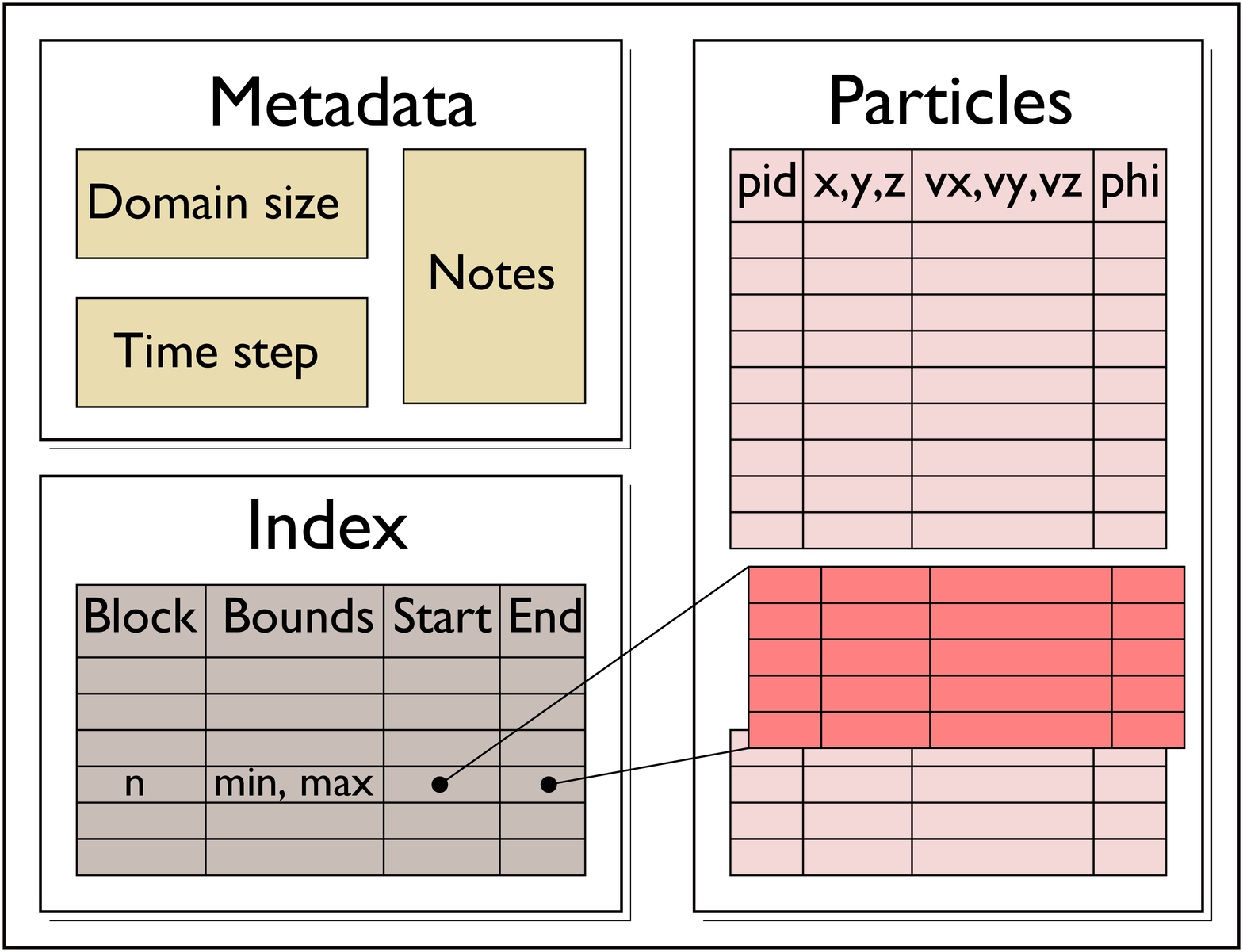}
  \caption{File schema for analysis output enables spatial queries of
    particle data in a high-level, portable, self-describing format.} 
  \label{fig:pnetcdf}
\end{figure}

\textit{Many processes per file.} The default I/O strategy used by
HACC, called GenericIO, partitions the processes in a system-specific
manner, and each partition writes data into a custom self-describing
file format. Each process writes its data into a distinct region of
the file in order to reduce contention for file-system-level page
locks. Within the region assigned to each process, each variable is
written contiguously. On modern supercomputers, such as IBM Blue Gene
and Cray X series machines, dedicated I/O nodes execute special I/O
forwarding system software on behalf of a set of compute nodes. For
example, on IBM Blue Gene/Q systems, one I/O node is assigned to 128
compute nodes. By writing one file per I/O node, the total number of
files is reduced by at least the ratio of compute nodes to I/O nodes,
and a high percentage of the peak available bandwidth can be captured
by the reading and writing processes. By partitioning the processes by
I/O-node assignment and providing each process with a disjoint data
region, we are taking advantage of the technique successfully used by
the GLEAN I/O library~(\citealt{glean}) on the Blue Gene/P and Blue
Gene/Q systems. The I/O implementation can use MPI I/O, in either
collective or non-collective mode, or (non-collective) POSIX-level I/O
routines.

Importantly, 64-bit cyclic-redundancy-check (CRC) codes are computed
for each variable for each rank, and this provides a way to validate
data integrity. This detects corruption that occurs while the data is
stored on disk, while files are being transferred in between systems,
and while being transmitted within the storage subsystem(s). During a
recent run which generated ~100~TB of checkpoint files, a single error
in a single variable from one writing process was detected during one
checkpoint-reading process, and we were able to roll-back to a
previous checkpoint and continue the simulation with valid
data. Furthermore, over the past two years, CRC validation has
detected corrupted data from malfunctioning memory hardware on storage
servers, misconfigured RAID controllers and bugs in (or miscompiled)
compression libraries. The probability of seeing corrupt data from any
of these conditions is small (even while they exist), and overall,
modern storage subsystems are highly reliable, but when writing and
reading many petabytes of data at many facilities the probability of
experiencing faults is still significant. The CRC64 code is in the
process of being transformed into an open-source project\footnote{The
  library is available at
  http://trac.alcf.anl.gov/projects/hpcrc64/.}.

An example of GenericIO performance under production conditions on the
Blue Gene/Q system Mira is given in Table~\ref{table:genio-perf}. In
tests, I/O performance very close to the peak achievable has been
recorded. Under production conditions, we still achieve about
two-thirds of the peak performance.

\textit{A single file using parallel netCDF.} When peak I/O
performance is not required, and the system's MPI I/O implementation
can deliver acceptable performance, we can make use of a
parallel-netCDF-based I/O implementation (\citealt{li03}). The netCDF
format is used by simulation codes from many different science
domains, and readers for netCDF have been integrated with many
visualization and analysis tools. Because netCDF has an established
user community, and will likely be supported into the foreseable
future, writing data into a netCDF-based format should make
distributing data generated by HACC to other outside groups easier
than if only custom file formats were supported.

The file schema that we developed for the parallel netCDF format is
shown in Figure \ref{fig:pnetcdf}. As the figure shows, particles are
organized and indexed according to the spatial subdomains (blocks) of
the simulation, one block per process. The spatial extents of each
block are also indexed.

Currently, three modes of reading the parallel netCDF files are
supported. First, given some number of processes that need not be the
same as the number of blocks, particles can be redistributed uniformly
among the new number of processes while being read
collectively. Second, the particles in a single block can be retrieved
given the block ID. Third, particles in a desired bounding box can be
retrieved. The queried bounding box need not match the extents of any
one block and particles overlapping several blocks may be retrieved in
this manner.

The performance for writing the parallel netCDF output is shown in
Table \ref{table:pnetcdf-perf}. These tests were run on Hopper (Cray
XE6) at the National Energy Research Scientific Computing Center
(NERSC) with a Lustre file system. Because the file system is shared
by all running jobs, performance results will vary due to resource
contention; the values in Table \ref{table:pnetcdf-perf} are the means
of four runs for each configuration.

To put these results in perspective, consider the last row of
Table~\ref{table:pnetcdf-perf}. The peak performance expected for the
number of object storage targets (OSTs, 128 in this case) is
approximately 26.7~GiB/s (1 GiB = $2^{30}$ = 1,073,741,824
bytes). This value represents ideal conditions of writing large
amounts of data to one file per OST. In contrast, our strategy uses
one shared file with collective I/O aggregation and a high-level
format with index data in addition to raw particle arrays. Even so, we
achieve 56\% of the peak ideal bandwidth.

\begin{table}[!t]
\centering
\caption{GenericIO Performance (production runs on IBM Blue Gene/Q)}
\begin{tabular}{b{0.5in} b{0.5in} b{0.5in} b{0.5in} b{0.3in}}
\hline
No. Particles & No. Processes & File size (GiB) & Write time (s) &
Write Bandwidth (GiB/s) \\ 
\hline
$1024^3$  & 512     & 43.8 & 22.0  & 1.90 \\
$3200^3$ & 16384  &   1332.4 & 99.0 & 12.88  \\
$10240^3$ &  262144 & 43821.6 & 380.5 & 109.9  \\
\hline
\end{tabular}
\label{table:genio-perf}
\end{table}

\begin{table}[!t]
\centering
\caption{Parallel NetCDF Performance (test runs on Cray XE6)}
\begin{tabular}{b{0.5in} b{0.5in} b{0.5in} b{0.5in} b{0.3in}}
\hline
No. Particles & No. Processes & File size (GiB) & Write time (s) &
Write Bandwidth (GiB/s) \\ 
\hline
$512^3$  & 64   & 4.6 & 3.54  & 1.30 \\
$1024^3$ & 512  & 36  & 14.34 & 2.51  \\
$2048^3$ & 4096 & 288 & 19.10 & 15.1  \\
\hline
\end{tabular}
\label{table:pnetcdf-perf}
\end{table}

\section{Short-Range Force: Architecture Specific Implementations}     
\label{sec:specs}
In this section we go over the choice of algorithms deployed as a
function of nodal architecture, as well as the corresponding
optimizations implemented so far -- performance optimization is a
continuous process.

\subsection{Non-Accelerated Systems: The RCB Tree}
\label{rcb}

In order to evaluate the short-range force on non-accelerated systems,
such as the Blue Gene/Q, HACC uses an RCB tree in conjunction with a
highly-tuned short-range polynomial force kernel, as has been
discussed in Section~\ref{srf}. 

An important consideration in this implementation is the tree-node
partitioning step, which is the most expensive part of the tree build.
The particle data is stored as a collection of arrays -- the so-called
structure-of-arrays format. There are three arrays for the three
spatial coordinates, three arrays for the velocity components, in
addition to arrays for mass, a particle identifier, etc.  Our
implementation in HACC divides the partitioning operation into three
phases. The first phase loops over the coordinate being used to divide
the particles, recording which particles will need to be swapped.
Next, these prerecorded swapping operations are performed on six of
the arrays. The remaining arrays are identically handled in the third
phase. Dividing the work in this way allows the Blue Gene/Q hardware
prefetcher to effectively hide the memory transfer latency during the
particle partitioning operation and reduces expensive read-after-write
dependencies.

The premise underlying the multilevel timestepping scheme
(Section~\ref{time-stepper}) is that particles in higher density
regions will require finer short-range time steps. A local density
estimate can be trivially extracted from the same RCB tree constructed
for evaluating the short-range interparticle forces. Each ``leaf
node'' in the RCB tree holds some number of particles, the bounding
box for those particles has already been computed, and a constant
density estimate is used for all particles within the leaf node's
bounding box. Because the bounding box, and thus the density estimate,
changes as the particles are moved, the timestepping level assigned to
each leaf node is fixed when the tree is constructed. This implies
that, after particles have been moved, the leaf-node bounding boxes
may overlap. The force-evaluation algorithm is insensitive to these
small overlaps, and the effect on the efficiency of the force
calculation is apparently negligible.

\begin{figure}[t]
\begin{center}
\includegraphics[width=80mm]{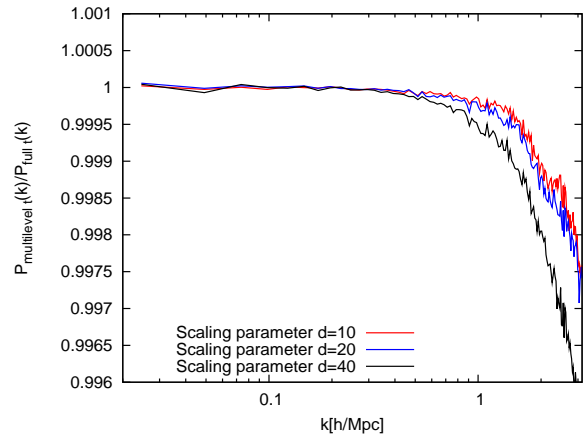}
\end{center}
\caption{Comparison of the multi-level time stepper with different
  values for $d$ [see Eq.~(\ref{kicks})] versus 5 sub-cycle steps for
  each particle. Shown is the ratio of the final power spectra for a
  small test problem (256~$h^{-1}$Mpc, 256$^3$ particles), going out
  to the particle Nyquist wave number. The multi-level time step
  approach leads to a speed-up of the full simulation by a factor of
  $\sim$2, with only a negligible change in the error.}
\label{time1}
\end{figure}

\begin{figure}[t]
\begin{center}
\includegraphics[width=70mm]{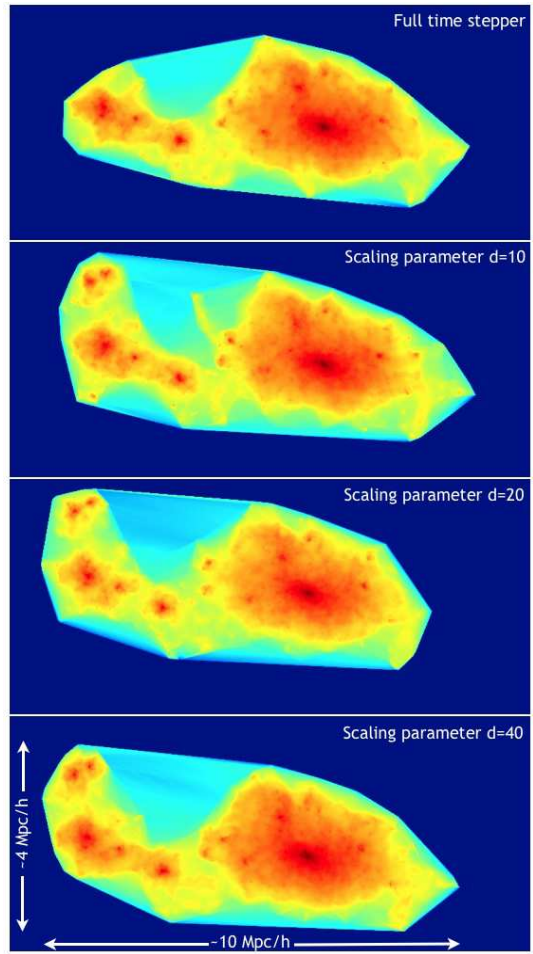}
\end{center}
\caption{Projected density field from a Delaunay
  tessellation-based density estimator of the largest (FOF, link
  length, $b=0.2$) halo at $z=0$ for the same run as in
  Figure~\ref{time1}, with different values for $d$. Minor variations
  in the outskirts of the halos are due to FOF linking `noise'.}
\label{time2}
\end{figure}

\begin{figure}[t]
\begin{center}
\includegraphics[width=80mm]{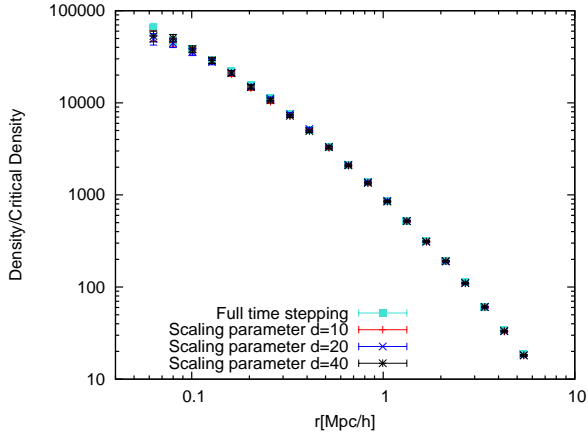}
\end{center}
\caption{Density profile for the halo depicted in Figure~\ref{time2}
  at corresponding values of the scaling parameter, $d$.}
\label{time2_dens}
\end{figure}

\begin{figure}[t]
\begin{center}
\includegraphics[width=75mm]{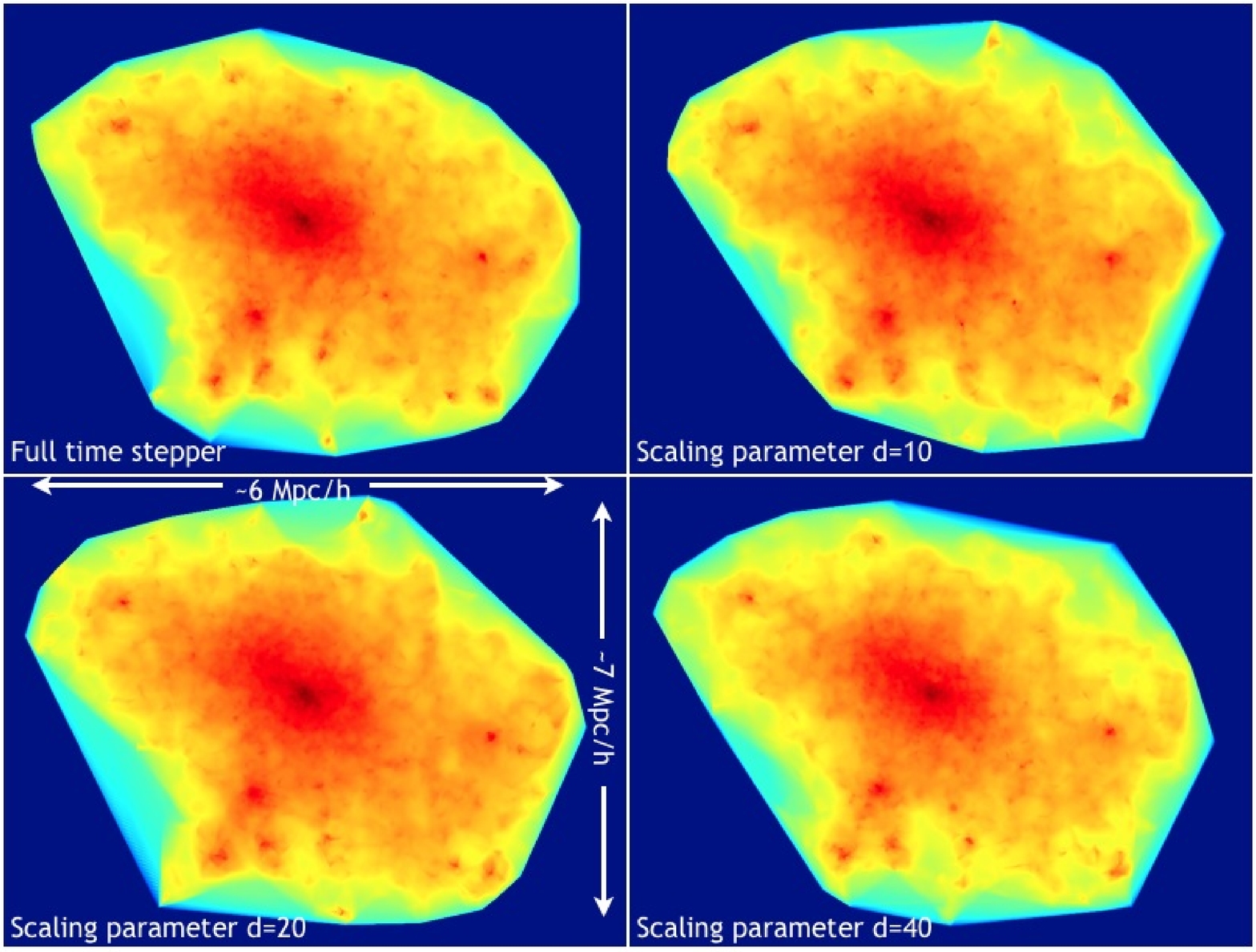}
\end{center}
\caption{Same as in Figure~\ref{time2} for the second largest halo.}
\label{time3}
\end{figure}

\begin{figure}[t]
\begin{center}
\includegraphics[width=80mm]{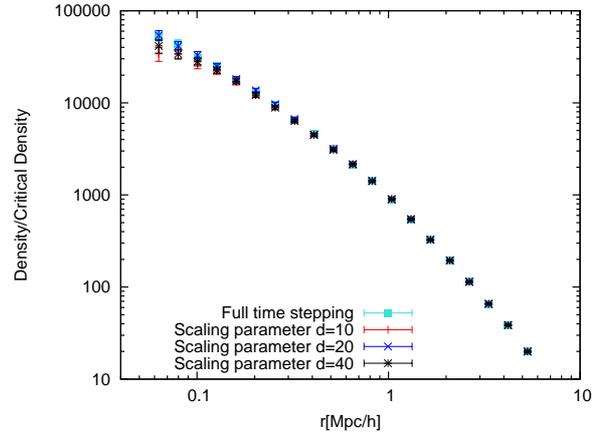}
\end{center}
\caption{Same as Figure~\ref{time2_dens} for the second-largest halo.} 
\label{time3_dens}
\end{figure}

In between consecutive long-range force calculations, each particle is
operated on by the kick (velocity update) and stream (position update)
operators of the short-range force. If a particle, based on the
density of its leaf node at the beginning of the subcycle, is evolved
using $n$ kicks, then it needs to be acted on by $2n$ stream operators
each evolving the particle by $dt/2n$. To ensure time synchronization,
these stream operators are further split such that all particles are
acted on by the same number of stream operators. The number of kicks
used for particles in a leaf node is determined by:
\begin{equation}\label{kicks}
  l = \left \lfloor \frac{\rho}{\rho_0} \right \rfloor /d + 1,
\end{equation}
where $d$ is an adjustable linear scaling parameter. In addition, the
maximum level is capped by an additional user-provided parameter.  We
show examples of accuracy control in the multi-level time stepping
scheme in Figures~\ref{time1} -- \ref{time3_dens}.

In Figure~\ref{time1}, we compare the power spectrum obtained from a
simulation with 5 sub-cycle steps for each particle with a result that
was obtained in the following way: 5 sub-cycles per step per particle
are used until $z=1$; since the clustering at that point is still
modest, this point is reached relatively quickly. After $z=1$ we
evolve each particle with at least 2 sub-cycles and allow -- depending
on the density -- two more levels of sub-cycling. In this test,
setting the scaling parameter $d$ to 20 or 10 leads to accurate
results, better than 0.2\% out to the particle Nyquist wavenumber,
$k_{\rm Ny}\equiv \pi/\Delta$, where $\Delta$ is the initial
inter-particle separation (the test case used 256$^3$ particles in a
256~$h^{-1}$Mpc volume with 500~PM steps.) In precision cosmology
applications, one desires better than $1\%$ accuracy, and this is
attained at wavenumbers less than $k_{\rm
  Ny}/2$~(\citealt{coyote1}). Consequently, the current error limits
are comfortably wihtin the required bounds, the error at $k_{\rm
  Ny}/2$ being only $\sim 0.05\%$. Using the multi-level stepping can
speed up the simulation by a factor of two (changing $d$ in the range
shown leaves the performance unaffected). We have carried out a suite
of convergence tests, concluding that the setting with $d=20$
satisfies our accuracy requirements.

At much smaller length scales, the power spectrum test above can be
augmented by checking the stability of small scale structures in the
halos as the adaptive time-stepping parameters are varied. As typical
examples thereof, we show results for the largest and second-largest
halos (identified using a `friends-of-friends' or FOF algorithm) in
the same simulation discussed above in Figures~\ref{time2} --
\ref{time3_dens}. The halo density field is computed via a
tessellation-based method in three dimensions and then projected onto
a two-dimensional grid; details of this implementation will be
presented elsewhere~(\citealt{rangel14}). The angle-averaged
(spherical) density profiles are also shown in
Figures~\ref{time2_dens} and \ref{time3_dens}. As can be seen from
these results, aside from trivial differences due to the FOF linking
noise, the halo substructure depends relatively mildly on the choice
of the values of the scaling parameter, $d$, over the chosen ranges
used.

Aside from optimizing the number of force evaluations, one also has to
minimize the time spent in evaluating the force kernel. This is a
function of the design of the compute nodes. Here we provide a
description of the Blue Gene/Q-specific short-range force kernel and
how it is optimized. (Very similar implementations were carried out
for Cray XE6 and XC30 systems.)  As mentioned earlier, the compactness
of the short-range interaction (Cf. Section~\ref{sec:lrange}), allows
the kernel to be represented as
\begin{equation}
f_{SR}(s)=(s+\epsilon)^{-3/2}-f_{grid}(s)
\label{force}
\end{equation}
where $s={\mathbf r}\cdot{\mathbf r}$, $f_{grid}(s)$ is a 5-th order
polynomial in $s$, and $\epsilon$ is a short-distance cutoff (Plummer
softening). This computation must be vectorized to attain high
performance; we do this by computing the force for every neighbor of
each particle at once. The list of neighbors is generated such that
each coordinate and the mass of each neighboring particle is
pre-generated into a contiguous array. This guarantees that 1) every
particle has an independent list of particles and can be processed
within a separate thread; and 2) every neighbor list can be accessed
with vector memory operations, because contiguity and alignment
restrictions are taken care of in advance. Every particle on a leaf
node shares the interaction list, therefore all particles have lists
of the same size, and the computational threads are automatically
balanced.

The filtering of $s$, i.e., checking the short-range condition, can be
processed during the generation of the neighbor list or during the
force evaluation itself; since the condition is likely violated only
in a number of ``corner'' cases, it is advantageous to include it into
the force evaluation in a form where ternary operators can be combined
to remove the need of storing a value during the force
computation. Each ternary operator can be implemented with the help of
the {\tt{fsel}} instruction, which also has a vector equivalent. Even
though these alterations introduce an (insignificant) increase in
instruction count, the entire force evaluation routine becomes fully
vectorizable.

There is significant flexibility in choosing the number of MPI ranks
versus the number of threads on an individual Blue Gene/Q
node. Because of the excellent performance of the memory sub-system, a
large number of OpenMP threads -- up to 32 per node -- can be run to
optimize performance. Concurrency in the short-range force evaluation
is exposed by, first, building a work queue of leaf-node-vs-tree
interactions, and second, executing those interactions in parallel
using OpenMP's dynamic scheduling capability. Each work item
incorporates both the interaction-list building and the force
calculation itself for each leaf node's particles.

To further increase the amount of parallel work, HACC builds multiple
RCB trees per rank. First, the particles are sorted into fixed bins,
where the linear size of each bin is roughly the scale of the
short-range force. An RCB tree is then constructed within each bin,
and because this process in each bin is independent of all other bins,
this is done in parallel. This parallelization of the tree-building
process provides a significant performance boost to the overall force
computation. When the force on the particles in each leaf node is
computed, not only must the parent tree be searched, but so must the
other 26 neighboring trees. Because of the limited range of the
short-range force, only nearest neighbors need to be considered. While
searching many neighboring trees adds extra expense, the trees are
individually not as deep, and so the resulting walks are less
expensive. Also, because we distribute `(leaf node, neighboring tree)'
pairs among the threads, this scheme also increases the amount of
available parallelism post-build (which helps with thread-level load
balancing). All told, using multiple trees in this fashion provides a
significant performance advantage over using one large tree for the
entire domain.

\subsection{Cell-Accelerated Systems}

The first version of HACC was originally written for the IBM
PowerXCell 8i-accelerated hardware of Roadrunner, the first machine to
break the Petaflop barrier. This architecture posed three critical
challenges, all of which continue to be faced in one way or the other
on all accelerated systems. A more detailed description of the Cell
implementation and the Roadrunner architecture is given in
\cite{hacc1}. (See also \citealt{sriram}.)

\begin{figure}[t]
  \centering \includegraphics[width=3.2in,angle=0]{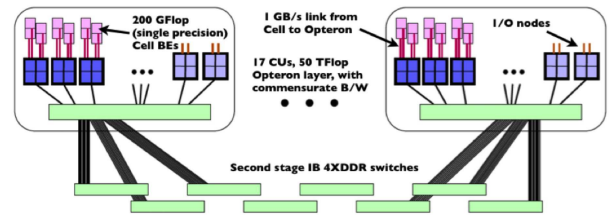}
  \caption{Roadrunner architecture schematic: The accelerator units
    (IBM Cell~BEs) are shown in pink, the CPUs (AMD Opterons) in blue,
    and the (fat tree) interconnect in green. Note that communication
    across the Cell~BE layer has to proceed indirectly, passing
    through the CPU layer.}
\label{rr_arch}
\end{figure}

The three challenges for a Roadrunner-style architecture are as
follows. (i) {\em Memory Balance.} The machine architecture
(Figure~\ref{rr_arch}) has a top layer of conventional multi-core
processors (in this case, two dual-core AMD Opterons) to which are
attached IBM PowerXCell 8i Cell Broadband Engines (Cell~BEs) via
an eight-lane PCI-E bus. The relative performance of the Opterons is small
compared to that of the Cell~BEs, by roughly a factor of 1:20, but
they carry half the memory and possess access to a communication
fabric that is balanced to their level of computational
performance. For large-scale N-body codes, memory is a key limiting
factor, therefore the code design must make the best use possible of
the CPU layer (this situation continues in current and future
accelerated systems, as discussed further below). (ii) {\em
  Communication Balance.} The Cell~BEs dominate the computational
resource, but are starved for communication, due to the relatively
slow PCI-E link to the host CPU (Figure~\ref{rr_arch}). From the point
of view of the compute/communication ratio, such a machine is 50-100
times out of balance. We note that this situation also continues to
hold in the current generation of accelerated systems such as CPU/GPU
or CPU/Xeon Phi machines: The computational approach taken must
therefore maximize computation for a given amount of data
motion. (iii) {\em Multiple Programming Models.} Accelerated systems
have a multi-layer programming model. On the CPU level, standard
languages and interfaces can be used (HACC uses C/C++/MPI/OpenMP) but
the accelerator units often have a hardware-specific programming
paradigm. (Although attempts to overcome this gap now exist, the
results -- in actual practice -- are not yet compelling.) For this
reason, it is desirable to keep the code on the accelerator (the
Cell~BE in this case) as simple as possible and avoid elaborate
coding. In addition, it also proved advantageous to keep the data
structures and communication patterns on the Cell~BE as
straightforward as possible to optimize computational efficiency.

With these challenges in mind, HACC is matched to the machine
architecture as follows: At the first level of code organization, the
medium/long range force is handled by the FFT-based method that
operates at the Opteron layer, as for all other architectures.  At
this layer, only grid information is stored and manipulated (except
when carrying out analysis steps). Particles live only at the Cell
layer. There is a rough memory balance between the grid and particle
data, matching well to the memory organization on the machine, and
combating the first challenge mentioned above. The particle-grid
deposition and grid-particle interpolation steps are performed at the
Cell layer with only grid information passing between the two
layers. This compensates for the limited bandwidth available between
the Cell~BEs and the Opterons. The local force calculations reside at
the Cell level. This addresses the second challenge, as aided by the
particle overloading discussed in Section~\ref{sec:overload}.  Because
implementing complicated data structures at the Cell level is
difficult, and conditionals are best avoided, our choice for the local
force solve is a direct particle-particle interaction. To make this
interaction more efficient, we use a chaining mesh to control the
number of interactions~(see, e.g., \citealt{hockney}). This leads to
an efficient, hardware-accelerated P$^3$M algorithm, thereby
overcoming the third challenge (\citealt{hacc2}).

\begin{figure}
 \centering \includegraphics[width=2.5in,angle=0]{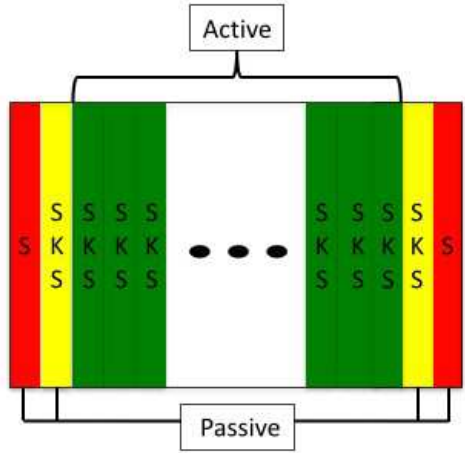}
 \caption{Figure illustrating the partitioning of active and passive
   particles. The outside (red) slabs of data are solely streamed, and
   the yellow and green slabs calculate the full short range (SKS)
   force. In analogy, the red and yellow slabs can be thought of as
   the overloaded `passive' particles, and the green slabs are `active',
   which are correctly updated.  The finest partitioning is achieved
   in 5 slab blocks, where there is only one (green) corrected slab
   updated. }
\label{data_part}
\end{figure}

\subsection{GPU-Accelerated Systems}

As already discussed above, some of the challenges for GPU-accelerated
systems are very similar in spirit to those for Cell-accelerated
systems. The low-level GPU programming model (OpenCL or CUDA) adds
another layer of complexity and the compute to communication balance
is heavily skewed towards computing. One major difference between the
two architectures is the memory balance: While on the Cell-accelerated
systems the Cell layer has the same amount of memory as the CPU layer,
this is generally not the case for CPU/GPU systems. For example, the
Cray XK7 system, Titan, at Oak Ridge, has 32~GB of host-side memory on
a single node, with only 6~GB of GPU memory. This adds yet another
challenge, that of memory imbalance. We have overcome this by
partitioning the local data into smaller overlapping blocks, which can
fit on device memory. Very similar in spirit to particle overloading,
the boundaries of the partitions are duplicated, such that each block
can be evolved independently. We emphasize again that the long/medium
range calculations on this architecture remain unchanged, and only the
short-range force kernel needs to be optimized.

The data partitioning is illustrated in Figure~\ref{data_part}. We
utilize a two-dimensional decomposition of data blocks, which are in
turn composed of slabs that are spatially separated by the extent of
the short-range force -- roughly 3 grid cells (see
Figure~\ref{ppforce}). In complete analogy to particle overloading,
the data blocks are composed of `active' particles (green) that are
updated utilizing the `passive' particles (yellow and red) from the
boundary. The (red) edge slabs are solely streamed, as opposed to
performing the full SKS time stepping described in
Section~\ref{time-stepper}. This mediates the error inflow on these
passive particles, as they can only ``see'' the interior particles
within the domain. We note that since the data has been decomposed
into smaller independent work items, these blocks can now be
communicated to any nodes that have the available resources to handle
the extra workload. Hence, this scheme provides for a straightforward
load-balancing algorithm by construction. Details of the error
analysis and load balancing schemes will be described in an upcoming
paper devoted to the GPU implementation of HACC~\citep{frontiere14}. 

\begin{figure}[t]
  \centering \includegraphics[width=3.2in,angle=0]{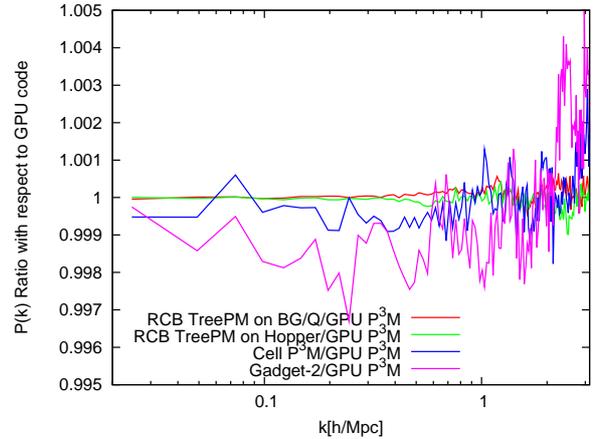}
  \caption{Ratio of the power spectra from four different runs with
    respect to the HACC GPU implementation, time-stepping conditions
    for all HACC versions held fixed.  The comparison shows the Blue
    Gene/Q PPTreePM version (red solid), the X86 TreePM version run on
    a Cray XE6 (green dashed), and the GPU P$^3$M version (blue
    short-dashed). We also show the comparison with a {\sc Gadget-2}
    run. The agreement is very good -- the TreePM runs agree with the
    GPU version of HACC to better than 0.1\% up to the particle
    Nyquist wavenumber. The level of agreement with {\sc Gadget-2} is
    noteworthy because it is a completely independent code. }
\label{comp_pk}
\end{figure}

\section{Code Verification and Testing}
\label{sec:verif}

\begin{figure*}[t]
\begin{center}
\includegraphics[width=40mm]{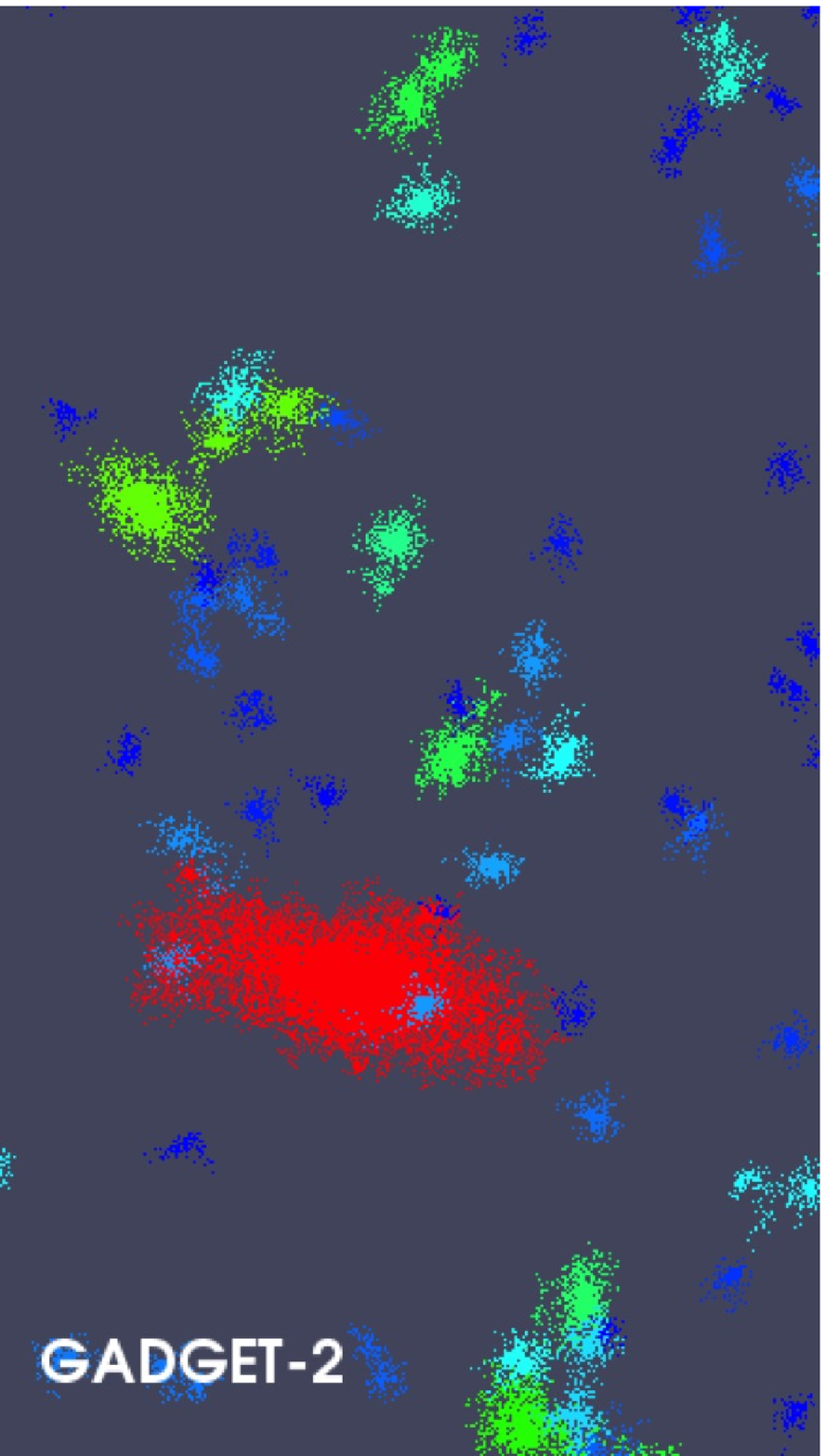}
\includegraphics[width=40mm]{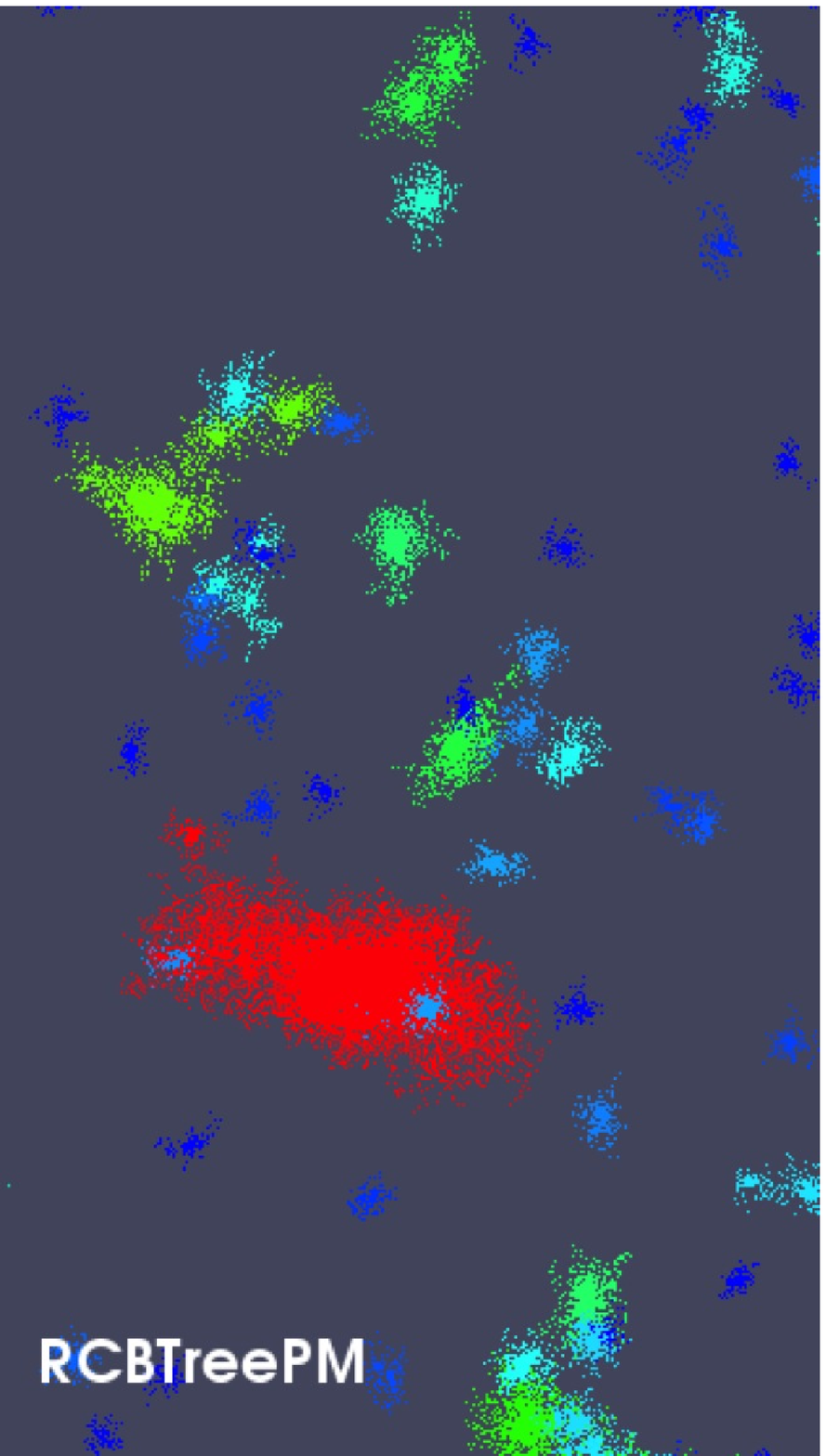}
\includegraphics[width=40mm]{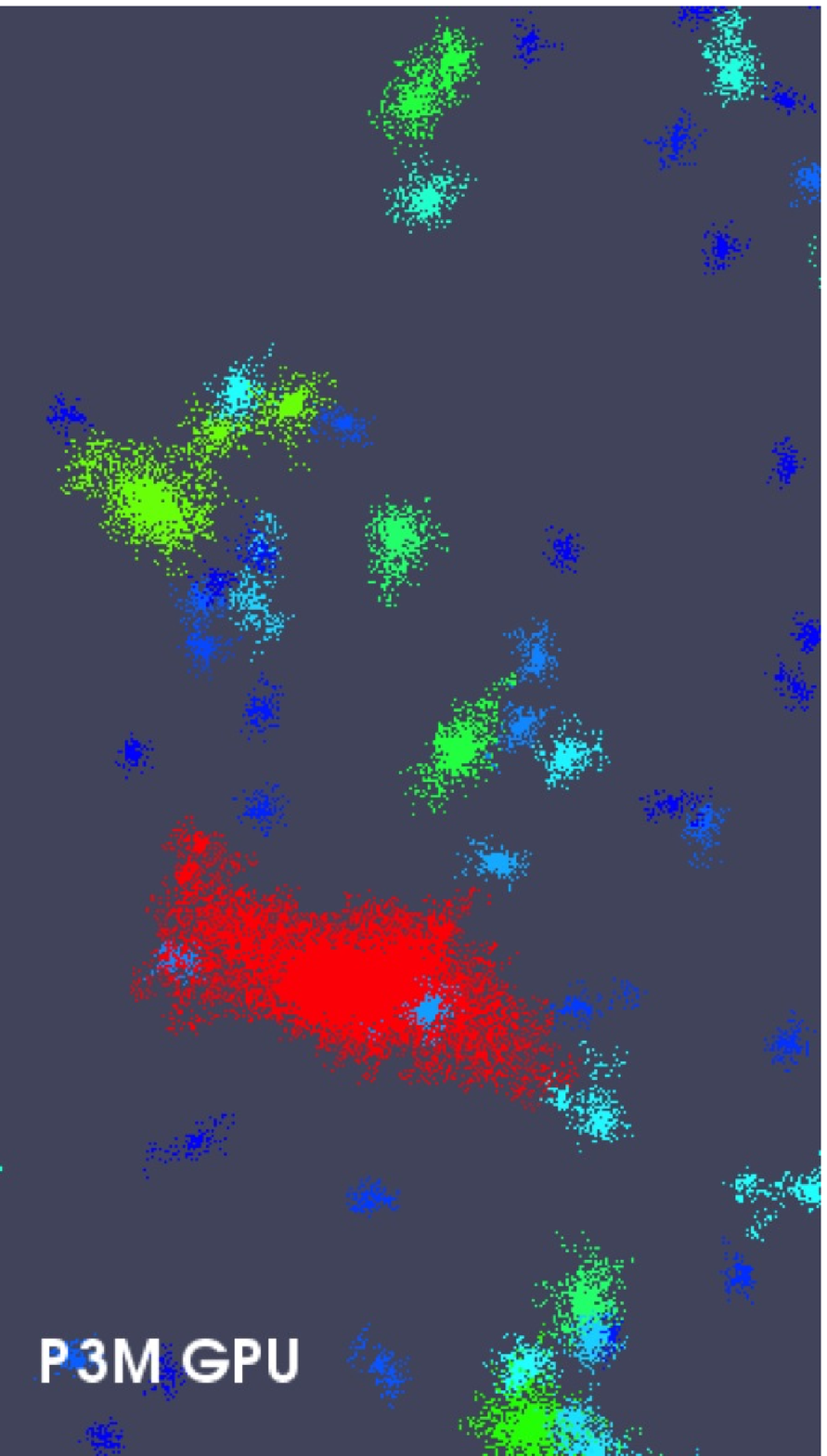}
\includegraphics[width=40mm]{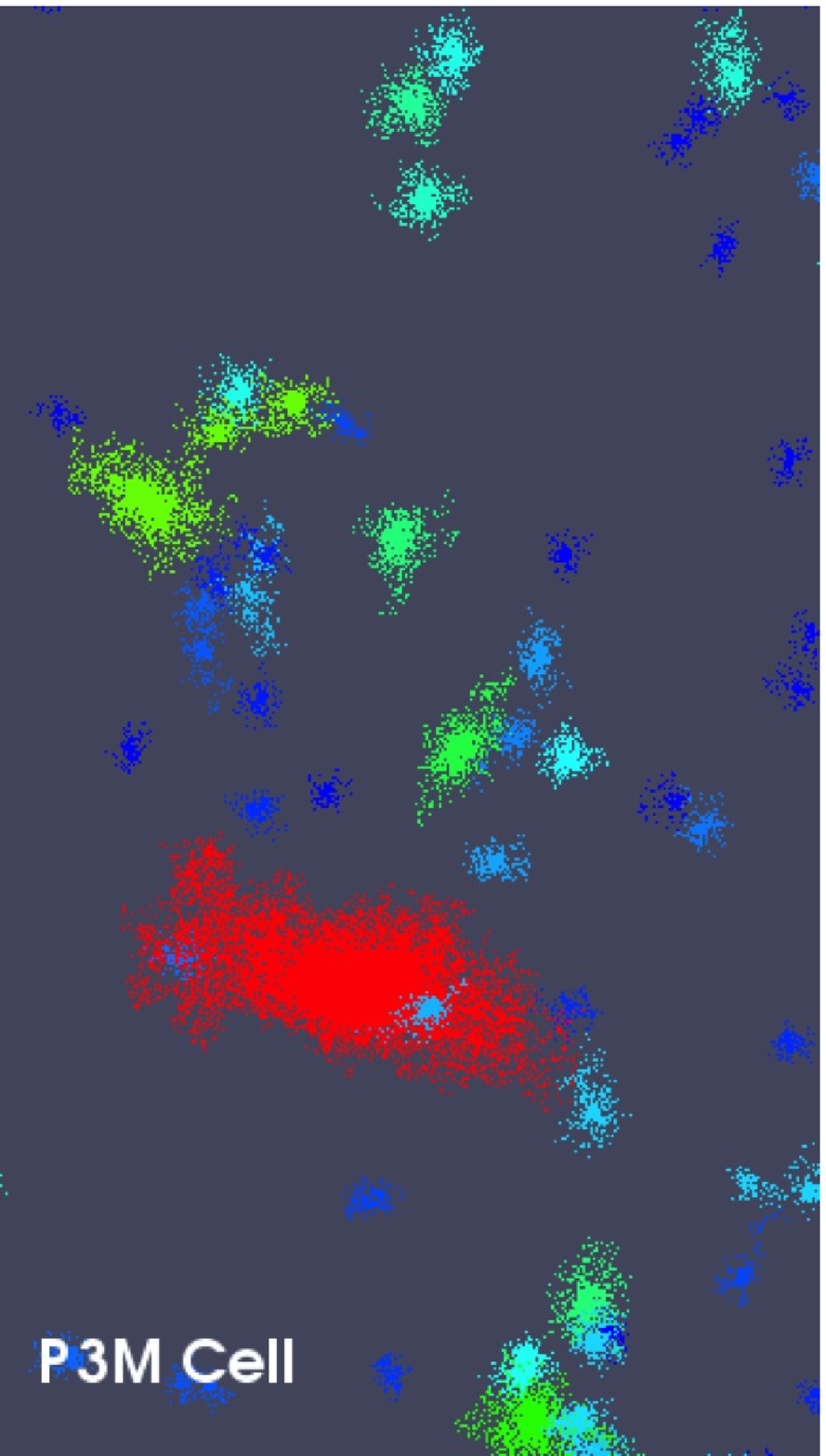}
\end{center}
\caption{Comparison of halo particles between {\sc Gadget-2} (left)
  and three HACC implementations, the PPTreePM version (Cray XE6), the
  GPU version, and the Cell version. The smallest halos shown consist
  of 100 particles (dark blue), the largest halos have up to
  $\sim$19,000 particles (red). Green colors show halos in the few
  thousand particle range. If a dark blue halo is missing in one of
  the images, this is due to the mass cut at 100 particles (the halo
  has fallen below a threshold, but actually exists). The linking
  length in the comparison was chosen to be $b=0.168$. The very good
  level of overall agreement is clearly evident. }
\label{comp_vis}
\end{figure*}

\begin{table*}
\begin{center} 
  \caption{Comparison of a sample of halo statistics from the code
    comparison runs, as extracted from the ParaView analysis. The
    total number of halos is N$_h$, the number of particles, N$_p$,
    the FOF link length, $b$, and the velocity dispersion is denoted by
    $\sigma_v$. \label{tab1}}
\vspace{0.3cm}
\begin{tabular}{c|cccc}
\hfill & {\sc Gadget-2} & RCBTreePM & P$^3$M-GPU & P$^3$M-Cell \\ 
\hline
N$_h$, $b=0.2$ & 9707 & 9638 & 9636 & 9634\\  
N$_h$, $b=0.168$ & 8817 & 8734 & 8732 & 8728\\
N$_p$, most massive halo, $b=0.2$ & 22,587 & 21,802 &
22,114 & 22,240\\ 
N$_p$, most massive halo, $b=0.168$ & 18,728 & 18,656 &
19,047 & 19,088 \\ 
range of $\sigma_v$ [km/s], $b=0.2$ & [132.4, 1109.9]&
[134.4, 1126.2] &[134.2, 1101.3]& [133.2, 1101.16]\\ 
range of $\sigma_v$ [km/s], $b=0.168$ & [133.5, 1144.5]&
[145.5, 1141.3] &[151.3, 1128.6]& [143.7, 1146.7]\\ 
\end{tabular}
\end{center}
\end{table*}

\begin{figure}[t]
\begin{center}
\includegraphics[width=80mm]{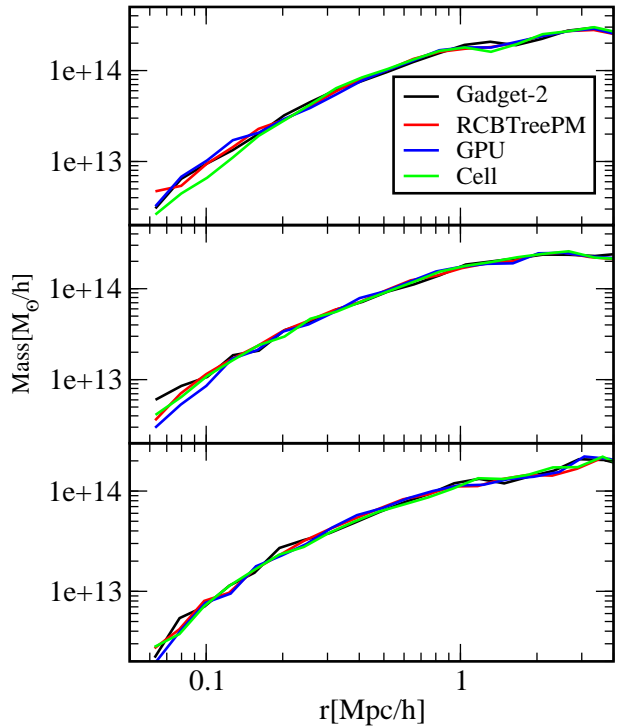}
\end{center}
\caption{Comparison of the mass profiles of the three biggest halos in
  the test simulation, showing close agreement within the binning shot
  noise scatter (larger at smaller radii). }
\label{comp_prof}
\end{figure}

\begin{figure}[t]
  \centering \includegraphics[width=80mm,angle=0]{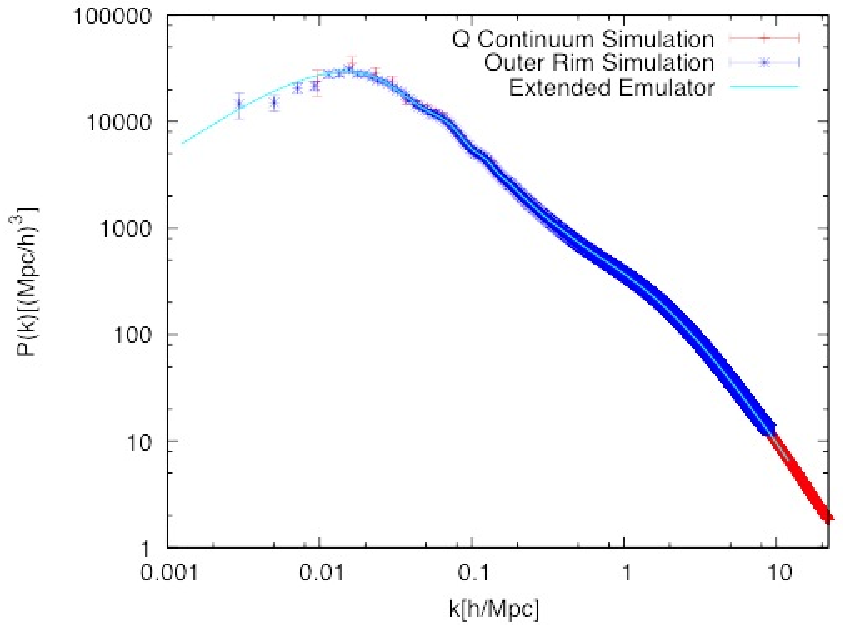}
  \caption{Matter fluctuation power spectra from the Outer Rim and Q
    Continuum simulations at $z=0$. The solid curve is the
    prediction from the extended Coyote emulator of
    \cite{heitmann14}. The agreement across the two runs is at the
    fraction of a percent level, while the agreement with the emulator
    is at the 2\% level, which is the estimated emulator accuracy. }
\label{big_pk}
\end{figure}

HACC has been subjected to a large number of standard convergence
tests (second-order time-stepping, halo profiles, power spectrum
measurements, etc.). In this section we focus mostly on a subset of
HACC test results using the setup of the code comparison project,
originally carried out in~\cite{heitmann05}. In that work, a set of
initial conditions was created for different problems (mainly
different volumes) and a number of cosmological codes were run on
those, all at their nominal default settings. The final outputs were
compared by measuring a variety of statistics, including matter
fluctuation power spectra, halo positions and profiles, and halo mass
functions. The initial conditions and final results from the tests are
publicly available and have been used subsequently by other groups for
code verification, e.g for {\sc Gadget-2}~\citep{gadget2}, and most
recently for Nyx~\citep{almgren}. In addition, we will also show some
results from recently carried out large-scale simulations.

We first restrict attention to the larger volume simulation
(256~$h^{-1}$Mpc) and compare HACC results with those found for {\sc
  Gadget-2}, as published in~\cite{gadget2}. While the simulation is
only modest in size (256$^3$ particles) it does present a relatively
sensitive challenge and is capable of detecting subtle errors in the
code under test. Not only are statistical measures such as the power
spectrum robust indicators of code accuracy, but visual inspection of
the particle data itself presents a quick qualitative check on code
behavior and correctness; it is particularly valuable in identifying
problems at early stages of code development. We use ParaView for this
purpose (\citealt{woodring11}).

The code comparison test was run with a force resolution of $\sim
7~h^{-1}$kpc, very similar to what was used in the {\sc Gadget-2}
simulation. We compare results from different HACC versions (PPTreePM
on the Blue Gene/Q and Hopper, P$^3$M on a Cell-accelerated and a
GPU-accelerated system) with those from {\sc Gadget-2}. The result for
the matter power spectrum is shown in Figure~\ref{comp_pk}, where, as
in Figure~\ref{time1}, we show results up to $k_{\rm Ny}$. All the
code results are very close to each other (note the scale on the
y-axis).  The agreement over the full $k$-range is better than 0.5\%,
including {\sc Gadget-2}. Considering the different implementations of
force solvers and time steppers, this closeness is very reassuring,
particularly as no effort was made to fine-tune the level of
agreement. The TreePM versions of HACC and the CPU/GPU version agree
to better than a tenth of a percent.

Next we present results from a more qualitative, but nevertheless,
very detailed comparison, shown in Figure~\ref{comp_vis}. In this
test, we identify all particles that belong to halos with at least 100
particles per halo.  This is done within ParaView, using an FOF halo
finder with linking length $b=0.168$.  The particles are colored with
respect to halo mass. The most massive halo in the image (colored in
red) has a mass of $\sim 1.6\cdot 10^{15}~h^{-1}$M$_\odot$. While
there are differences in the images -- as is to be expected -- the
overall agreement is striking. Almost all small halos exist in all
images (the ones that are missing are just below the cut of 100
particles, but they do actually exist) and many of the fine details
within the halo structures of the larger halos are well-matched.

The mass profiles of the three largest halos in the simulation are
shown in Figure~\ref{comp_prof}. The binning shot noise dominates the
comparison at small radii, but beyond that the agreement is very
good. Other quantitative halo comparison statistics are given in
Table~\ref{tab1}, to further illustrate the close match of the results
from all of the different algorithms.

We illustrate the dynamic range and accuracy of the HACC approach by
comparing results for the matter power spectrum from the `Outer Rim'
and `Q Continuum' runs in Figure~\ref{big_pk}. The Q Continuum run on
Titan had $\sim 550$ billion particles, a 1.3~Gpc box, with a mass
resolution, $m_p\sim 10^8$~M$_\odot$, and the Outer Rim run on Mira
had $\sim 1.1$ trillion particles, a 4.225~Gpc box, with a mass
resolution, $m_p\sim 10^9$~M$_\odot$. The numerical results (run with
different short-range force algorithms) agree to fractions of a
percent, and agree at the 2\% level with the (extended) Coyote
emulator predictions of \cite{heitmann14}, which is at the level of
accuracy expected from the emulator.

\begin{figure}[t]
\begin{center}
\includegraphics[width=85mm]{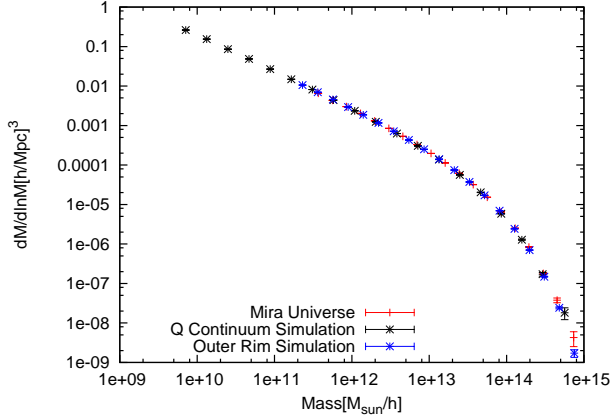}
\end{center}
\caption{The (FOF, $b=0.168$) halo mass function for three large
  simulations run with HACC, measured at $z=1$. The box sizes range
  from $1-4$~Gpc and the particle number in each simulation ranges
  from $\sim 30$ billion to over one trillion (mass resolutions range
  from $\sim 10^8$~M$_\odot$ to $\sim 10^{10}$~M$_\odot$; for details,
  see text).}
\label{mf_big}
\end{figure}

Finally, we show results for the halo mass function at very large
simulation scales illustrating excellent agreement across different
sized simulations carried out using different short-range force
implementations. Figure~\ref{mf_big} shows the (FOF, $b=0.168$) halo
mass function at $z=1$ resulting from three different simulations with
the same cosmology, (i) a run from the Mira Universe suite ($\sim 30$
billion particles, 2.1~Gpc box, mass resolution, $m_p\sim
10^{10}$~M$_\odot$), (ii) the Q Continuum run on Titan, and (iii) the
Outer Rim run on Mira.

\section{In situ Analysis Tools}
\label{sec:tools}

\begin{figure}[!t]
  \centering
  \includegraphics[width=3.0in]{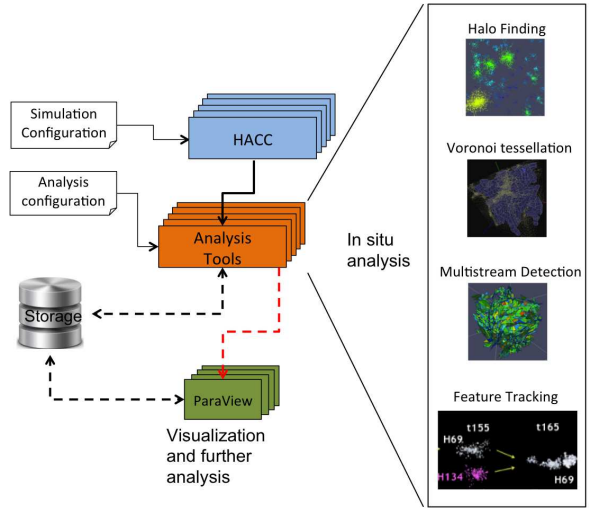}
  \caption{The {\em in situ} analysis framework provides the ability to
    apply various analysis tools and methods, e.g., halo finders,
    multistream diagnostics, feature tracking (halo merger trees), and
    Voronoi tessellation, and connects to run-time or postprocessing
    visualization tools, all while the simulation is running.}
  \label{fig:framework}
\end{figure}

An entire suite of {\em in situ} analysis tools (CosmoTools) for HACC is
under continuous development, driven by evolving science goals;
CosmoTools exists in both {\em in situ} and stand-alone versions (to be used
for off-line processing). The overall structure of the {\em in situ}
framework is shown in Figure \ref{fig:framework}. Output from the
analysis is available either at run time or for post-processing. In
the former case, a ParaView server can be launched and connected to
the simulation through a tool called Catalyst. In the latter case,
results of the {\em in situ} analysis are written to a parallel file system
and later retrieved.

\begin{figure*}[t]
\centering \includegraphics[width=3.22in,angle=0]{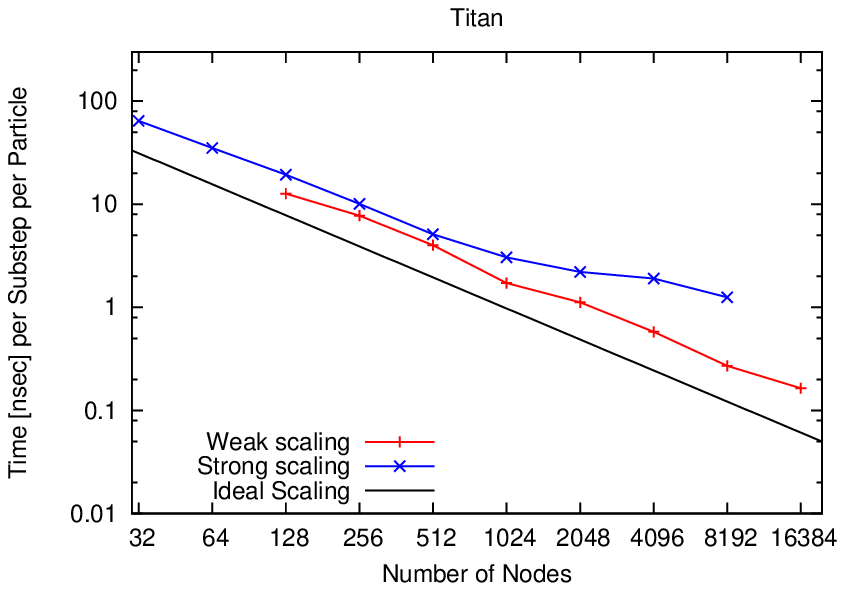}  
\centering \includegraphics[width=3.22in,angle=0]{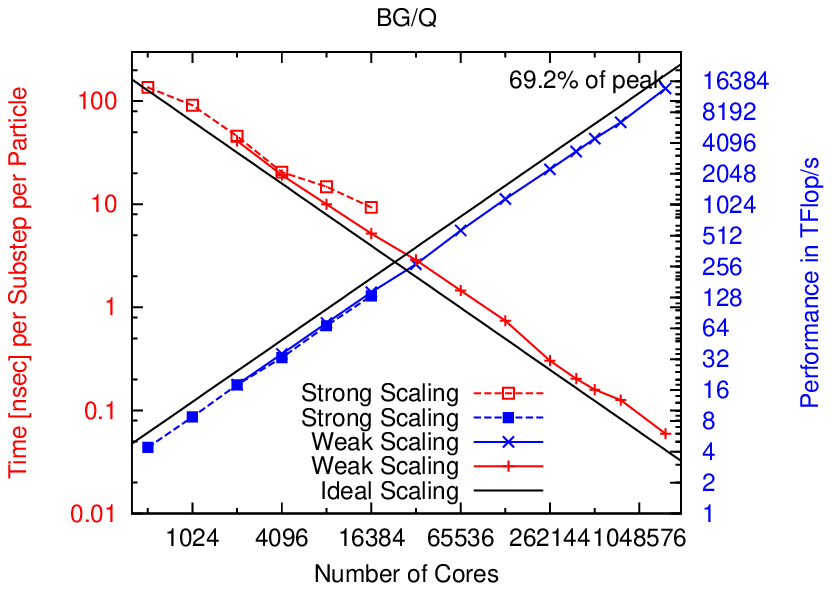}
\vspace{-0.6cm}

\caption{\small{\em Early weak and strong scaling results on Titan
    (left panel) and from Blue Gene/Q systems (right panel, results
    from Mira and Sequoia).  Weak scaling results are reported for a
    constant number of particles and physical volume per node/core
    (see text for details on both systems).  Strong scaling results
    are for a fixed-size problem -- 1024$^3$ particles in a
    1~$h^{-1}$Gpc box. Optimal scaling is represented by the solid
    black lines.}}
\label{comp}
\end{figure*}

Several standard tools for the analysis of cosmological simulations
are part of the {\em in situ} framework, such as
halo finders~(\citealt{woodring11}) and merger tree constructors.  The
first tool to be part of this framework that works on the full
particle data to produce field information is a parallel Voronoi
tessellation that computes a polyhedral mesh whose cell volume is
inversely proportional to the distance between particles. Such a mesh
representation acts as a continuous density field that affords
accurate sampling of both high- and low-density regions. Connected
components of cells above or below a certain density can also
approximate large-scale structures. Two important criteria for {\em in situ}
analysis filters are that they should scale similarly as the
simulation and have minimal memory overhead. The parallel tessellation
approach meets these criteria; full details are given in
\cite{peterka_sc12}. The various tools can be turned on through the
configuration file for HACC, and the frequency of their execution is
also adjustable.

\section{Selected Performance Results}
\label{sec:perf}

HACC runs on a variety of supercomputing platforms and has scaled to
the maximum size of some of the fastest machines in the world,
including Roadrunner at Los Alamos, Sequoia at Livermore, Titan at Oak
Ridge, Mira at Argonne, and Hopper at NERSC. We have carried out
detailed scaling and performance studies on the Blue Gene/Q
systems~(\citealt{hacc3}) and on Titan; a sample of our results is
presented below.

For both systems, we carried out weak and strong scaling tests. For
the weak scaling tests we fix a physical volume and number of
particles per node. When scaling up to more nodes, the volume and
particle loading therefore increases, while the mass resolution stays
constant. The wall-clock time for a run should hence stay constant if
the code scales or, equivalently, the time to solution per particle
per step should decrease. The absolute performance measured in TFlops
per seconds will rise while the percentage per peak will stay
constant. For our weak scaling tests, the particle mass is $\sim
5\cdot 10^{10}$M$_\odot$ and the force resolution, 6~kpc. All
simulations are for a $\Lambda$CDM model. Simulations of cosmological
surveys focus on large problem sizes, therefore the weak scaling
properties are of primary interest.

For the weak scaling test on the Blue Gene/Q systems, we ran with 2
million particles in a $\sim$(100~$h^{-1}$Mpc)$^3$ volume per core,
using a typical particle loading in actual large-scale simulations on
these systems.  Tests with 4 million particles per core produce very
similar results.  As demonstrated in Figure~\ref{comp} (right panel),
weak scaling is ideal up to 1,572,864 cores (96 racks, all of
Sequioa), where HACC attains a peak performance of 13.94~PFlops and
parallel efficiency of $90\%$.  The largest test simulation on Sequoia
evolved $\sim$3.6 trillion particles and a (very high accuracy)
particle substep took $\sim 0.06$~ns for the full high-resolution
code. The scaling results were obtained by averaging over 50
sub-cycles.

On Titan we ran with 32 million particles per node in a fixed (nodal)
physical volume of $(256~h^{-1}{\rm Mpc})^3$, representative of the
particle loading in actual large-scale runs (the GPU version was run
with one MPI rank per node).  The results are shown in the left panel
of Figure~\ref{comp}. In addition (not shown) we have timing results
for a 1.1~trillion particle run, where we have kept the volume per
node the same but increased the number of particles per node by a
factor of two to 64.5 million. As for the Blue Gene/Q systems, HACC
weak-scales essentially perfectly up to the full machine.

For the strong scaling tests, we chose the same problem size on both
systems, a (1000 $h^{-1}$Mpc)$^3$ volume with 1024$^3$ particles. This
is a rather small problem and strong scaling is expected to break down
at some point. The results for both systems are shown in
Figure~\ref{comp} -- the strong scaling regime is remarkably large. On
the Blue Gene/Q system, we demonstrate strong scaling between 512 and
16384 cores (with somewhat degraded performance at the largest
scale). For Titan, we increase the number of nodes for this problem
from 32 to 8192, almost half of the machine. As can be seen in the
left panel in Figure~\ref{comp}, strong scaling only starts degrading
after 2048 nodes. (The results for Titan have been improved since
these earlier tests, they will be reported in \citealt{frontiere14}). The
significance of the strong scaling tests is in showing how well a code
can perform as the effective memory per core reduces -- HACC can run
very effectively at values as low as 100MB/core, which is roughly
equivalent to the memory per core available in next-generation
many-core systems.

\section{Conclusion and Outlook}
\label{sec:conc}

The impressive scale and quality of data from sky survey observations
requires a correspondingly strong response in theory and modeling. It
is widely recognized that large-scale computing must play an essential
role in shaping this response, not only in interpreting the data, but
also in optimizing survey strategies and validating data and analysis
pipelines. The HACC simulation framework is designed to produce
synthetic catalogs and to run large campaigns for precision
predictions of cosmological observables.

The evolution of HACC continues to proceed in two broad directions:
(i) the further development of algorithms (and their optimization) for
future-looking supercomputing architectures, including areas such as
power management, fault-tolerance, exploitation of local non-volatile
random-access memory (NVRAM) and investigation of alternative
programming models, and (ii) addition of new physics capabilities in
both modeling and simulation (e.g., gas physics, feedback processes,
etc.) and in the analysis part of the framework (e.g., increased
sophistication in semi-analytic galaxy modeling, and an associated
validation program).

The use of HACC for large-scale simulation campaigns covers
applications such as those required to construct cosmological
emulators (\citealt{HHHN, HHHNW, coyote3, kwan13a, kwan13b,
  heitmann14}), to determine covariance matrices (\citealt{sunayama}),
to help optimize survey design and test associated pipelines with
synthetic catalogs (\citealt{mw14}), and, finally, to carry out
MCMC-based parameter estimation across multiple cosmological probes
(\citealt{higdon}). A major component of the future use of HACC is
therefore related to the production and exploitation of large
simulation databases, including public access and analysis; work in
this area is in progress with a number of collaborators.

\section{Acknowledgments}
For useful interactions and discussions over time, the authors thank
Jim Ahrens, Viktor Decyk, Nehal Desai, Wu Feng, Chung-Hsing Hsu, Rob
Latham, Pat McCormick, Rob Ross, Robert Ryne, Paul Sathre, Sergei
Shandarin, Volker Springel, Joachim Stadel, and Martin White. We
acknowledge early contributions to HACC by Nehal Desai and Paul
Sathre. Running on a number of supercomputers, a large fraction of
which were in their acceptance phase, required the generous assistance
and support of many people. We record our indebtedness to Susan
Coghlan, Kalyan Kumaran, Joe Insley, Ray Loy, Paul Messina, Mike
Papka, Paul Rich, Adam Scovel, Tisha Stacey, William Scullin, Rick
Stevens, and Tim Williams (Argonne National Laboratory), Brian Carnes,
Kim Cupps, David Fox, and Michel McCoy (Lawrence Livermore National
Laboratory), Lee Ankeny, Sam Gutierrez, Scott Pakin, Sriram
Swaminarayan, Andy White, and Cornell Wright (Los Alamos National
Laboratory), and Bronson Messer and Jack Wells (Oak Ridge National
Laboratory).

Argonne National Laboratory's work was supported under U.S. Department
of Energy contract DE-AC02-06CH11357.  Part of this research was
supported by the DOE under contract W-7405-ENG-36. Partial support for
this work was provided by the Scientific Discovery through Advanced
Computing (SciDAC) program funded by the U.S. Department of Energy,
Office of Science, jointly by Advanced Scientific Computing Research
and High Energy Physics. This research used resources of the ALCF,
which is supported by DOE/SC under contract DE-AC02-06CH11357 and
resources of the OLCF, which is supported by DOE/SC under contract
DE-AC05-00OR22725. Some of the results presented here result from
awards of computer time provided by the Innovative and Novel
Computational Impact on Theory and Experiment (INCITE) and ASCR
Leadership Computing Challenge (ALCC) programs at ALCF and OLCF.

\end{document}